\documentclass[a4paper]{aa}
\usepackage{natbib}
\bibpunct{(}{)}{;}{a}{}{,}
\usepackage{amssymb}
\usepackage{graphicx}
\usepackage{euscript}

\usepackage{epsfig}
\usepackage{caption}
\begin{document}

\title{The effects of vertical outflows on disk dynamos}
\author{A. Bardou \inst{1} \and B. von Rekowski \inst{1} \and W. Dobler \inst{1}
        \and A. Brandenburg \inst{1,2} \and A. Shukurov \inst{1}}

\institute{Department of Mathematics, University of Newcastle, 
Newcastle upon Tyne NE1 7RU, United Kingdom 
\and Nordita, Blegdamsvej 17, DK2100 Copenhagen, Denmark}

\offprints{A. Bardou, \email{Anne.Bardou@ncl.ac.uk}}

\date{Received ??; accepted ??, 2001}

\abstract{
We consider the effect of vertical outflows on the mean-field dynamo in a thin 
disk. These outflows could be due to winds or magnetic buoyancy.
We analyse both two-dimensional finite-difference numerical solutions 
of the axisymmetric dynamo equations and a free-decay
mode expansion using the thin-disk approximation. Contrary to expectations, 
a vertical velocity can enhance
dynamo action, provided the velocity is not too strong. In the nonlinear regime this
can lead to super-exponential growth of the magnetic field.
\keywords{Accretion,accretion disks -- Magnetic fields -- Galaxies: magnetic fields}
}

\maketitle

\section{Introduction}
Magnetic fields play an important role in accretion disks. Inside the disk,
for example, they allow the magneto-rotational instability to develop \citep{BH98}. 
This instability
can lead to the development of turbulence which in turn transports angular momentum
radially outwards, thus allowing the matter to accrete. On the other hand, large scale
magnetic fields are often considered necessary for launching winds or jets from 
accretion disks.
The origin of such large scale magnetic fields is still unclear: they 
may be advected from the surrounding medium or be generated by a dynamo
inside the disk. Advection appears unlikely as turbulence leads to enhanced 
viscosity and magnetic diffusivity, so that the two are of the same order 
\citep[Prandtl number of order unity;][]{PFL76}. 
Thus, turbulent magnetic diffusivity
can compensate the dragging of the field by viscously induced accretion flow 
\citep{vBa89,LPP94,HPB96}. Dynamo action is a plausible mechanism 
for producing magnetic fields in accretion disks \citep{Pud81,SL88}. However, 
dynamo magnetic
fields can affect winds from accretion disks \citep{CPA98,Cam99,BDS00}.
On the other hand, the wind can also affect the dynamo. In particular, an 
enhancement by wind was 
suggested by \citet{Bra93} in the context of galactic dynamos.
In the same context, \citet{EGR95} also consider the effects of shear 
in the wind.

In this paper, we consider the generation of magnetic field by a dynamo acting 
in a thin
accretion disk. We provide a study of the effects of vertical velocities. 
We show how vertical velocities can enhance the dynamo,
allowing a larger growth rate and leading to super-exponential growth of the
magnetic field in the nonlinear regime. 

The vertical velocities considered can have several origins.
They can, for example, be due to a wind emanating 
from the disk or to magnetic buoyancy inside the disk.
Note that we are interested here in magnetic buoyancy as generating vertical
outflows. We neglect the associated dynamo effect \citep{MSS99}.

The paper is organized as follows. In Sect.~2, we describe the two
complementary approaches used and specify our choice of parameters. In 
Sect.~3, we
discuss the linear results without vertical outflows. Results with
vertical outflows are discussed in Sect.~4 for the linear regime
and in Sect.~5 for the nonlinear regime. Sect.~6 presents our
conclusions.

\section{The models}
Simulations of dynamo-generated turbulence in accretion disks have shown that, if
stratification  is included, a large scale field can be generated
whose structure and dependence on boundary conditions closely resemble 
what is known from mean-field $\alpha\omega$-dynamo theory \citep{Bra98}.
It has been suggested that nonlinear effects can suppress the $\alpha$-effect
to insignificant values \citep{VC92}, but as recent simulations have 
shown, this does not affect  the very possibility of mean-field
dynamo action, but merely the time-scale on which the large scale field
is established \citep{Bra01}.
With these justifications in mind, we adopt the mean-field approximation. 
The evolution of the magnetic field is given by 
\begin{equation}
\frac{\partial \vec{B}}{\partial t} = \vec{\nabla} \times (\vec{V} \times \vec{B}
                  + \alpha \vec{B} - \eta \vec{\nabla} \times \vec{B}),
\label{eq1}
\end{equation}
where $\eta$ is the turbulent magnetic diffusivity and $\alpha$ the coefficient
quantifying the $\alpha$-effect.
We ignore anisotropies of $\alpha$
because only the toroidal diagonal component of the $\alpha$-tensor 
matters in the so-called $\alpha\Omega$-regime. The antisymmetric part of the
$\alpha$-tensor corresponds to turbulent diamagnetism 
which tends to push magnetic fields into the disk,
and therefore just lowers the effective wind speed.
Furthermore, as shown by \citet[ Fig.~6]{GSS96}, turbulent diamagnetism is 
relatively unimportant for dynamo numbers less than 200, and our results mostly refer to this range. 
Given the substantial uncertainties in the mean-field transport coefficients,
we decided to adopt the simplest possible approach. In order to minimize the
number of parameters in our model,
we also neglect the radial component of the velocity and thus our velocity field is
$\vec{V} = (0,r \Omega,V_z)$ in cylindrical coordinates ($r, \varphi, z$) (see comments in
Sect.~2.2 on the consequences of this assumption).

We use two approaches to solve this equation.
Firstly, a 2D numerical simulation based on a finite-difference scheme is applied
(Sect.~2.1). Secondly, we develop a 1D free-decay mode expansion (Sect.~2.2). Details of 
these two complementary approaches are given below. 
Each of these two methods has its own advantages. The finite-difference simulation allows
investigation of the nonlinear behaviour and applies to 2D geometry. On the other hand, the
free-decay mode expansion gives the full eigenstructure and provides a tool for
qualitative analysis. 
Furthermore, boundary conditions corresponding to an ideal vacuum
can be implemented easily in the 1D model, whereas our 2D model
where the disk is embedded in a halo can provide only an approximation to a vacuum outside
the disk. Our two approaches are therefore complementary.

The problem is governed by three dimensionless numbers which appear 
in the dimensionless form of Eq.~(\ref{eq1}).
Let $h$, the half-thickness of the disk, be a characteristic length-scale, 
$\alpha_0$ a characteristic value of $\alpha$,
$V_{z0}$ a characteristic vertical velocity and $\eta_0$ a characteristic 
value of $\eta$.  We also define the rotational shear 
$S(r) = r \mathrm{d}\Omega / \mathrm{d} r$. The dimensionless numbers are the 
azimuthal magnetic Reynolds 
number
\begin{equation}
R_\omega = \frac{S h^2}{\eta_0},
\label{R1}
\end{equation}
the magnetic Reynolds number based on the $\alpha$-effect
\begin{equation}
R_\alpha = \frac{\alpha_0 h}{\eta_0},
\label{R2}
\end{equation}
and the vertical magnetic Reynolds number
\begin{equation}
R_v = \frac{V_{z0} h}{\eta_0} . 
\label{R3}
\end{equation}
Note that $\alpha_0$ and $S$ depend on $r$ and thus the magnetic Reynolds numbers
$R_\omega$ and $R_\alpha$ also depend on $r$. However, we assume later that $\alpha_0 S$ is 
approximately constant.
In this work, we adopt the $\alpha\omega$-approximation for the dynamo. 
In that case, for thin disks, the growth rates depend on the dynamo number
$\cal D \equiv R_\alpha R_\omega$ only and 
not on the individual 
values of $R_\alpha$
and $R_\omega$. However, note that the ratio of the radial to the azimuthal 
magnetic field $|B_r| / |B_\varphi|$ is about 
$\sqrt{|R_\alpha| / |R_\omega|}$ \citep{RSS88}.
Moreover, \citet{Pud81} showed that the dynamo number $\cal D$ is
approximately constant in a thin accretion disk.

\subsection{The finite-difference model}
To ensure that the magnetic field $\vec{B}$ is solenoidal,
we evolve the induction equation in terms of the vector potential
$\vec{A}$, where $\vec{B} = \vec{\nabla} \times \vec{A}$,
\begin{equation}
{\partial \vec{A} \over \partial t} = \vec{V} \times \vec{B} 
                                      + \alpha \vec{B} - \eta \mu_0 \vec{j},
\label{eqA}
\end{equation}
where $\vec{j} = \vec{\nabla} \times \vec{B} / \mu_0$ is the current density
due to the mean magnetic field and $\mu_0$ is the magnetic permeability.
The $\alpha$-term $\alpha \vec{B}$ together with the velocity term
$\vec{V} \times \vec{B}$ drives the mean-field dynamo.
We restrict ourselves to axisymmetric solutions of Eq.~(\ref{eqA}). 
We do not make the thin-disk approximation in this model but
we consider thin disks by choosing an appropriate aspect ratio (see below).
We do not make 
the $\alpha\omega$-dynamo assumption explicitly i.e. we retain the $\alpha$-effect in
the equations for $A_r$ and $A_z$, but we restrict ourselves
to the $\alpha\omega$-dynamo regime by our choice of magnetic Reynolds
numbers $|R_\omega| / |R_\alpha| = 10^4 \mbox{--} 10^6$.

Our computational domain contains the disk embedded in a halo:
$0 \le r \le L_r{=}2$, $|z| \le L_z{=}1$, where the disk occupies the region
$0 \le r \le 1.5$, $|z| \le 0.15$, $R=1.5$ being the outer radius of the disk and
$h=0.15$ its half-thickness. This results in a disk aspect ratio of
$h/R = 0.1$ and the upper boundary of the domain at $z \simeq 7 h$.
Using a resolution of $101 \times 101$ grid points and a uniform grid,
the resulting mesh spacing is $\delta r = \delta z = 0.02$.
Equation~(\ref{eqA}) is solved using a sixth order finite-difference scheme in space
and a third order Runge--Kutta time advance scheme.

We control the symmetry of our solutions (dipolar or quadrupolar) by using
appropriate initial conditions.
On the boundaries of the computational domain (i.e. far away from the disk)
we impose the conditions
that the normal component of $\vec{A}$, $A_\perp$, vanishes together with the normal
derivative of the tangential components, $\vec{A}_\parallel$, 
\begin{equation}
A_\perp = 0, \quad {\partial \over \partial n} \vec{A}_\parallel = 0.
\end{equation}
This is similar to the ``normal-field'' condition, where the tangential
components of the magnetic field and the normal component of the
electric field $\vec{E}$ vanish, $\vec{B}_\parallel = \vec{0}$, $E_\perp = 0$.
To verify that the boundary conditions do not affect our results, we
have also tried perfectly conducting boundary conditions and found
differences in the magnetic field structure only in the close vicinity
of the boundaries.

The $\alpha$-coefficient $\alpha(r,z)$ is antisymmetric about 
the midplane and vanishes outside the disk. We adopt
\begin{equation}
\alpha(r,z) =
\left\{ \begin{array}{cl}
        \alpha_0(r) \sin(\pi z / h) & 
	\mbox{for $|z| \leq h$}, \\
	0 &
	\mbox{for $|z| > h$}.
	\end{array} \right.
\label{alp}
\end{equation}
The radial profile of $\alpha(r,z)$ is smoothly cut off at $r > R$ and
at $r_i = 0.15$ where the rotational shear is very strong.

As appropriate for accretion disks, we adopt a softened Keplerian angular
velocity profile in $r$ in the whole
computational domain (including the halo as well as the disk), assuming
$\Omega$ to be $z$-independent:
\begin{equation}
\Omega(r) = \sqrt{G M \over r^3}
\left[ 1 + \left( {r_0 \over r} \right)^n \right]^{-{(n+1) / 2n}},
\end{equation}
where $G$ is Newton's gravitational constant, $M$ the mass of the central object, 
$r_0 = 0.05$ the softening
radius, and $n=5$. At $r = 0$, $\Omega$ vanishes as $r^{3/2}$.

The turbulent magnetic diffusivity is given by an interpolation between
$\eta_{\rm disk}$
in the disk midplane and $\eta_{\rm halo}$ in the halo, 
\begin{equation}
\eta(r,z) = \eta_{\rm halo} + (\eta_{\rm disk} - \eta_{\rm halo}) \xi(r,z).
\end{equation}
Here, the profile $\xi(r,z)$ defines the disk,
\begin{equation}
\xi(r,z) = g(r-r_0,d) \, g(|z| - z_0,d),
\end{equation}
where $g(x,d)$ is a smoothed Heaviside step function with a smoothing half-width
$d$ set to 8 grid zones. So $\xi(r,z)$ is equal to unity deep in the disk
and vanishes in the halo.

%We fix the magnetic diffusivity of the disk to $\eta_{\rm disk} = 1.5 \times 10^{-3}$ and
%consider two values of the magnetic diffusivity of the halo, $\eta_{\rm halo}$, 
%as described in Sect.~2.3.

To obtain the magnetic Reynolds numbers of Eqs.~(\ref{R1})--(\ref{R3})
we choose $\eta_0 = \eta_{\rm disk}$.

In accretion disks, the dynamo number is approximately constant
with radius. In this model this is achieved by setting
\begin{equation}
\alpha_0(r) = {{\cal D} \eta_{\rm disk}^2 \over S(r) h^3}.
\end{equation}

\subsection{The free-decay mode expansion}
The lowest order approximation for Eq.~(\ref{eq1}) in a thin disk results from retaining
derivatives with respect to $z$ alone.
In the $\alpha\omega$-approximation and with a constant $\eta$, Eq.~(\ref{eq1}) then becomes 
(see \citet{RSS88} for a form with $V_z = 0$) 
\begin{eqnarray}
\frac{\partial B_r}{\partial t} &=& 
  -\frac{\partial}{\partial z} (\alpha B_{\varphi}) 
+ \eta \frac{\partial^2 B_r}{\partial z^2}
- \frac{\partial}{\partial z}(V_z B_r), \label{eq2} \\
\frac{\partial B_{\varphi}}{\partial t} &=&
S B_r 
+ \eta \frac{\partial^2 B_{\varphi}}{\partial z^2}
- \frac{\partial}{\partial z}(V_z B_{\varphi}), \label{eq3}
\end{eqnarray}
where once again $S(r) = r d\Omega / dr$ and where $B_r$ and $B_\varphi$ are 
respectively the radial and azimuthal components of the magnetic field. 
Note that we have neglected $V_r$, the radial component of the
velocity in the disk, but the inclusion of this extra term would
be straightforward \citep{MSS00}.

We consider solutions with exponential time dependence
\begin{eqnarray}
B_r(z,t) &=& \widetilde{B_r}(z) \; \exp(\gamma t), \label{eq4}\\
B_{\varphi}(z,t) &=& \widetilde{B_{\varphi}}(z) \; \exp(\gamma t), \label{eq5}
\end{eqnarray} 
where $\gamma$ is the complex growth rate.
Measuring $z$ in units of the half-thickness $h$ of the disk, $t$ in units of 
the diffusion time $h^2/\eta$, the coefficient
$\alpha$ in units of a characteristic value $\alpha_0$ and $V_z$ in units of a 
characteristic value $V_{z0}$, the dimensionless form of
Eq.~(\ref{eq2}) and Eq.~(\ref{eq3}) is given by
\begin{eqnarray}
\gamma \widetilde{B_r} 
&=& - R_{\alpha} \frac{\mathrm{d}}{\mathrm{d} z} (\alpha \widetilde{B_{\varphi}})
  + \frac{\mathrm{d}^2 \widetilde{B_r}}{\mathrm{d} z^2}
- R_v \frac{\mathrm{d}}{\mathrm{d} z}(V_z \widetilde{B_r}), \label{base1}\\
\gamma \widetilde{B_{\varphi}} 
&=& R_{\omega} \widetilde{B_r} 
  + \frac{\mathrm{d}^2 \widetilde{B_{\varphi}}}{\mathrm{d} z^2}
- R_v \frac{\mathrm{d}}{\mathrm{d} z}(V_z \widetilde{B_{\varphi}}), \label{base2}
\end{eqnarray}
where we have kept the same notation for the dimensionless variables and
where $R_{\omega}$, $R_{\alpha}$ and $R_v$ are given by Eqs.~(\ref{R1}), (\ref{R2}) and (\ref{R3})
with $\eta_0 = \eta$.

We expand $\widetilde{B_r}$ and $\widetilde{B_{\varphi}}$ with respect to the 
orthonormal base formed
by the free-decay modes (Galerkin expansion),
\begin{eqnarray}
\widetilde{B_r} = \sum_{n=0}^{N} a_n B_r^{0,n} \: , \qquad 
\widetilde{B_{\varphi}} = \sum_{n=0}^{N} b_n B_{\varphi}^{0,n},
\label{eq9} 
\end{eqnarray}
where $a_n$ and $b_n$ are complex coefficients.
The free-decay modes are the solutions of the equation
\begin{equation}
\gamma_n^0 \vec{B}^{0,n} = \frac{\mathrm{d}^2}{\mathrm{d} z^2} \vec{B}^{0,n}.
\end{equation}
Vacuum boundary conditions are imposed at the disk surface. We impose them for each
free-decay mode so that $\widetilde{B_r}$ and $\widetilde{B_{\varphi}}$, and thus $B_r$ and
$B_{\varphi}$, satisfy these conditions
\begin{equation}
B_r^{0,n}|_{z=1} = B_{\varphi}^{0,n}|_{z=1} = 0.
\end{equation}
At the midplane, the boundary conditions are connected to the symmetry of the mode considered.
For quadrupolar modes we have
\begin{equation}
\left.\frac{\mathrm d B_r^{0,n}}{\mathrm d z}\right|_{z=0} 
= \left.\frac{\mathrm d B_{\varphi}^{0,n}}{\mathrm d z}\right|_{z=0} = 0,
\end{equation}
and for dipolar modes we have
\begin{equation}
B_r^{0,n}|_{z=0} = B_{\varphi}^{0,n}|_{z=0} = 0.
\end{equation}
Thus, the normalized eigenfunctions of free decay and the corresponding growth rates are, 
for quadrupolar symmetry
\begin{eqnarray}
B_{r,\varphi}^{0,n} &=& 
   \sqrt{2} \cos \left(\left({\textstyle{1 \over 2}} + n\right)\pi z\right) 
\:\:\:\: \mathrm{with} \:\: n= 0, 1 \ldots N, \\
\gamma_n^0 &=& - \left({\textstyle{1 \over 2}} + n \right)^2 \pi^2,
\end{eqnarray}
and for dipolar symmetry
\begin{eqnarray}
B_{r,\varphi}^{0,n} &=& \sqrt{2} \sin (n \pi z) \:\:\:\: \mathrm{with} \:\: 
n= 0, 1 \ldots N,\label{eqdipol1}\\
\gamma_n^0 &=& - n^2 \pi^2. \label{eqdipol2} 
\end{eqnarray}
Note that the free-decay dipolar mode $n=0$ would correspond to a uniform steady vertical 
magnetic field.
The free-decay modes are normalized such that
\begin{equation}
\int_0^1{\left(B_{r,\varphi}^{0,n}\right)^2} \mathrm dz = 1
\end{equation}
(except for the dipolar mode with $n=0$).
The free-decay modes are degenerate, forming doublets 
$(B_r^{0,n},0),(0,B_\varphi^{0,n})$ decaying at the same rate $\gamma_n^0$.

After making this expansion, the differential equations become an algebraic set of
equations forming an eigenvalue problem of $2(N+1)$ equations
\begin{eqnarray}
(\gamma - \gamma_m^0) a_m &=& - R_{\alpha} \sum_{n=0}^N b_n Y_{nm} -R_v \sum_{n=0}^N a_n W_{nm}^r,\\
(\gamma - \gamma_m^0) b_m &=& R_{\omega} a_m - R_v \sum_{n=0}^N b_n W_{nm}^{\varphi},
\end{eqnarray}
with $m = 0,1 \ldots N$.

The coefficients $W_{nm}^r$, $W_{nm}^{\varphi}$ and $Y_{nm}$ are defined by
\begin{eqnarray}
W_{nm}^r &=& \int_0^1 B_r^{0,m} (V_z \; B_r^{0,n})' \, \mathrm d z ,\\
W_{nm}^{\varphi} &=& \int_0^1 B_{\varphi}^{0,m} (V_z \; B_{\varphi}^{0,n})' \, \mathrm d z ,\\
Y_{nm} &=& \int_0^1 B_r^{0,m} (\alpha B_{\varphi}^{0,n})' \, \mathrm d z, 
\end{eqnarray}
where the prime denotes the derivative with respect to $z$.
We normalize the eigenfunctions (\ref{eq9}) such that 
\begin{equation}
\int\limits_0^1 \left[\left( \frac{\widetilde{B_r}}{R_\alpha}\right)^2 
\!\!\!\!+ \widetilde{B_\varphi}^2\right] \mathrm dz =
\sum_{n = 0}^N \left[ \left(\frac{|a_n|}{R_\alpha}\right)^2 
\!\!\!\!+ |b_n|^2 \right] = 1.
\end{equation}

In this model, the truncation level needed
depends on the values of $R_v$ and $\cal D$.
Typically, $N=4$ is enough to compute the eigenvalues (growth rates) of the 
problem whereas $N=9$ is usually required to compute the eigenfunctions. The
precision of all the results presented here was checked.

\subsection{Choice of parameters}
In this section, we discuss properties that are common to the two models.

The rotation of the thin disk is assumed to be close
to Keplerian
and independent of $z$ for the finite-difference model (see Sect.~2.1). Only the assumption
of the independence of $\Omega$ on $z$ is needed in the free-decay mode expansion.

We use two configurations: one where the magnetic diffusivity is the same in the disk and in 
the halo around the disk ($\eta_{\mathrm{halo}} = \eta_{\mathrm{disk}}$) and one 
with a low conductivity in the halo or a vacuum around 
the disk. Note that assuming a vacuum or low conductivity in the
halo above the disk is a rather good approximation for galaxies \citep{SS90,PSS93}. It also
allows for a connection with established theory. Note finally that the low-conductivity 
approximation ($\vec{j} \approx \vec{0}$) is a good approximation for potential fields 
($\vec{j} = \vec{0}$) 
and hence for a particular case of force-free fields ($\vec{j} \times \vec{B} = \vec{0}$).

The boundary conditions of the free-decay mode expansion imply 
a vacuum above the disk.
Technically, to achieve $\eta_{\mathrm{halo}} = \eta_{\mathrm{disk}}$ in this 
model, we restrict the dynamo action to an inner layer of the disk. 
The outer part of the disk is thus playing the role of the halo. The results
simply need to be rescaled for a proper comparison.

In the finite-difference model, we study the case 
$\eta_{\mathrm{halo}} = \eta_{\mathrm{disk}}$ and the case of a low conductivity 
halo with $\eta_{\mathrm{halo}} / \eta_{\mathrm{disk}} = 20$.
Note that, with this model, to approximate a vacuum in the halo we have
to choose $\eta_{\mathrm{halo}} / \eta_{\mathrm{disk}}$ as high as possible.
But, this choice is restricted because the higher this
ratio, the smaller the time step.

In this paper, the model with $\eta_{\mathrm{halo}} = \eta_{\mathrm{disk}}$
is referred to as the ``disk+halo'' model whereas the model with a vacuum or 
low conductivity 
above the disk is referred to as the ``disk+vacuum'' model.

In both models, the profile for $\alpha$ is taken to be
\begin{equation}
\alpha = \alpha_0 \sin\left(\pi \frac{z}{h}\right) \:\: \mbox{for} \:\: |z| \leq h
\end{equation}
(see Eq.~(\ref{alp}) for details concerning the finite-difference model), 
except in Fig.~\ref{fig0} where we take a linear
dependence of $\alpha$ with $z$ to compare with previous work.
The profile for $V_z$ is linear in $z$,
\begin{equation}
V_z = V_{z0} \frac{z}{h}.
\end{equation}

The accuracy of the models depends on the number of grid points in $r$ and $z$ for 
the finite-difference
model and on the level of truncation for the free-decay mode expansion. 

In this work, the two control parameters are the dynamo number $\cal D$ and
the vertical magnetic Reynolds number $R_v$. For reference, we give here possible
values for these two parameters. While dynamo numbers are rather low
in galactic disks, $|{\cal D}| \lesssim 10$ \citep{RSS88}, in accretion disks,
they can be as high as $|{\cal D}| \simeq 10^2-10^4$, depending on the Mach
number of the turbulence, see e.g. \citet{Pud81}. The value of
$R_v$ obviously depends on the origin of the vertical outflow. With magnetic 
buoyancy, a rough estimate can be obtained assuming that the vertical motions
occur at the Alfv\'en speed. Estimating the Alfv\'en speed from magnetic
equipartition, one gets a vertical magnetic Reynolds number $R_v$ of order
unity. 

In the following, the figures obtained with the finite-difference model
are labelled with ``FD'' and the ones obtained with the free-decay mode
(Galerkin) expansion with ``GE''.

\section{Linear results for $R_v = 0$}
In this section, we present results with no vertical velocity in the disk ($R_v = 0$).
Such a study is an interesting test for the quality of the results of the
free-decay mode expansion. It also gives a point of comparison between the numerical
simulation and the free-decay mode expansion. 

\begin{figure}
\centering
\includegraphics[height=7.5cm,width=8.5cm]{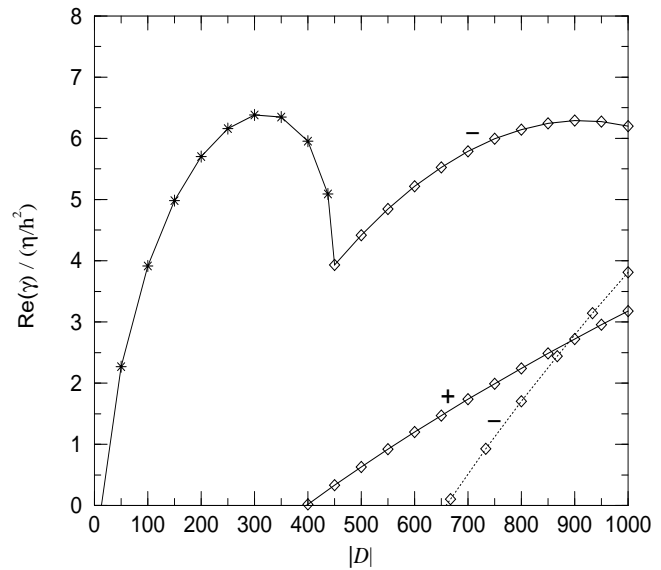}
\caption{Real part of the growth rate of the magnetic field as a function 
of $|{\cal D}|$,
with $\alpha = \alpha_0 z/h$, obtained with the free-decay-mode expansion
with $N \geq 9$. 
Asterisks denote non-oscillatory solutions
whereas diamonds denote oscillatory solutions.
Solid lines are for quadrupolar modes and dotted lines for dipolar modes. 
Signs above the curves indicate the sign of ${\cal D}$.}
\label{fig0}
\end{figure}

We first compare the results of the free-decay mode expansion (with $R_v = 0$) with the results
published by \citet{SL91}. We thus take a linear dependence of 
$\alpha$ on $z$, $\alpha = \alpha_0 z/h$. The calculated growth rates are plotted
in Fig.~\ref{fig0}. The agreement is good but a discrepancy is observed at 
large $|\cal D|$
where the increase of the growth rate with $|\cal D|$ is faster according 
to Stepinski \& Levy
than in our free-decay mode expansion for both quadrupolar and dipolar symmetry
(${\cal D}<0$). 
This discrepancy seems to be due to an insufficient number of
grid points in Stepinski \& Levy's model. Indeed, if we increase the truncation level
of the free-decay mode expansion, no change is observed in the results whereas if we decrease
the truncation level, we recover the results of Stepinski \& Levy. Note that the results
discussed in this paper are for $|{\cal D}| < 400$ where the agreement is excellent.

\begin{figure*}
\centering
\includegraphics[angle=0,height=7.5cm,width=8.5cm]{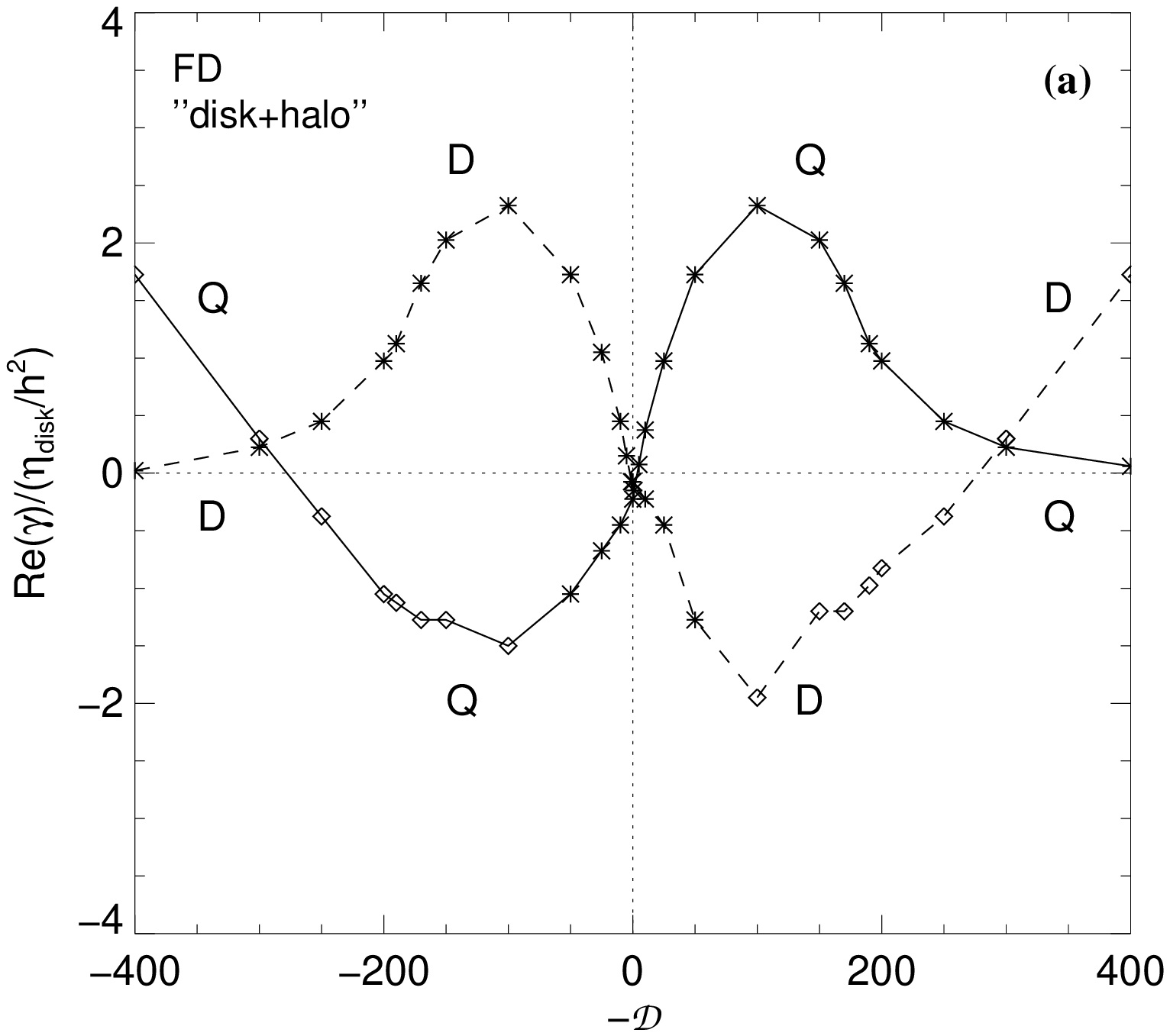}
\includegraphics[height=7.5cm,width=8.5cm]{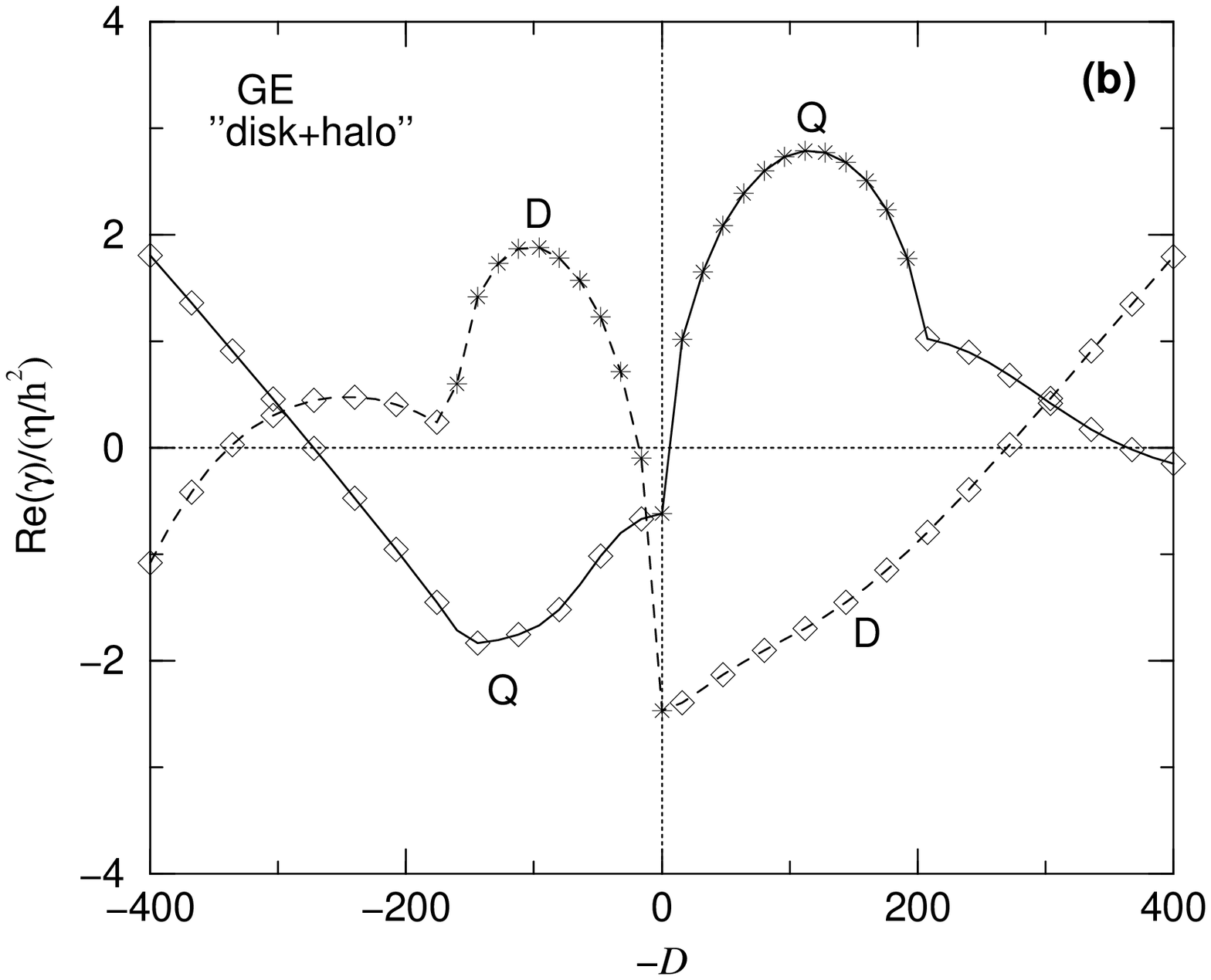}
\includegraphics[angle=0,height=7.5cm,width=8.5cm]{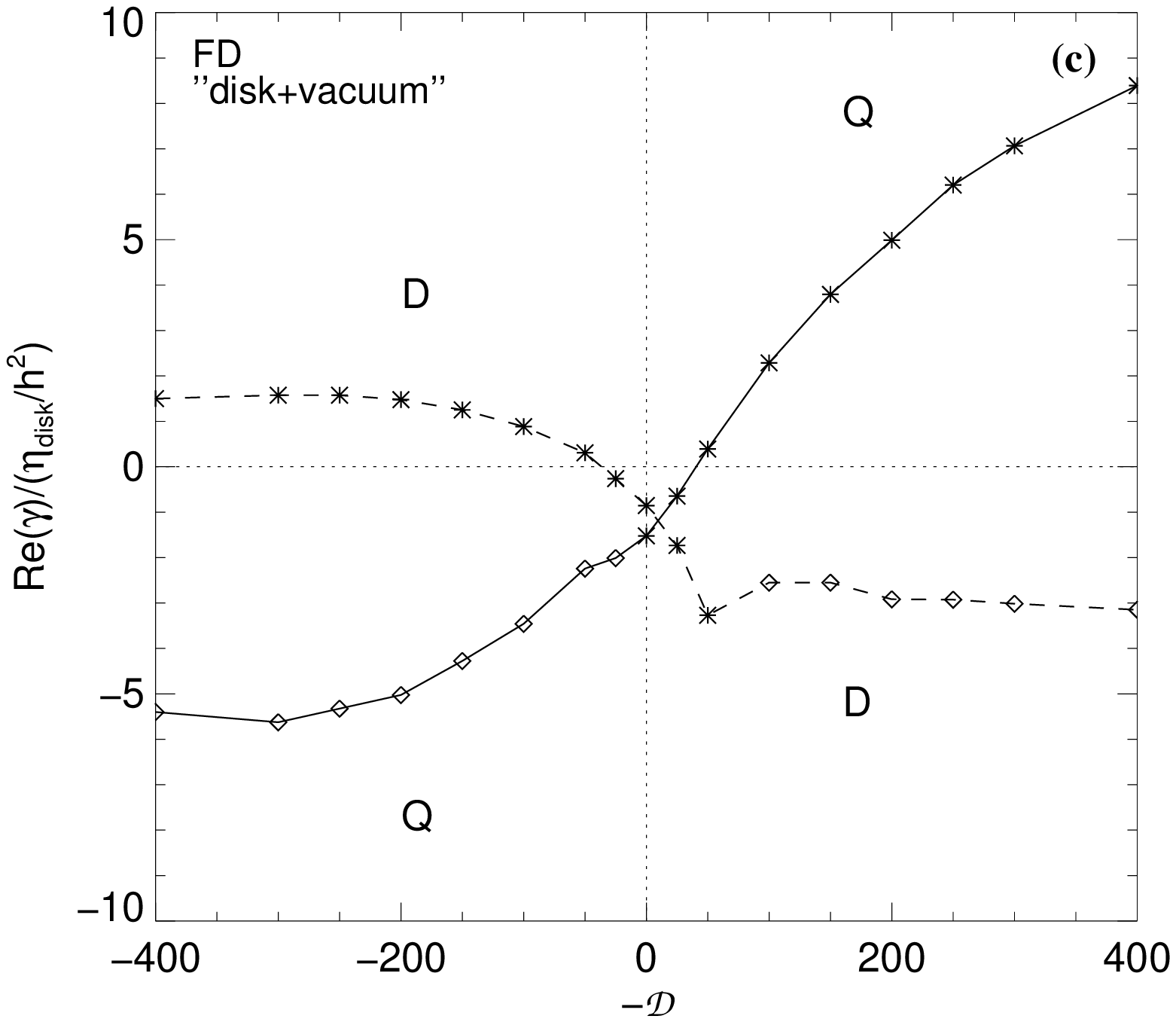}
\includegraphics[height=7.5cm,width=8.5cm]{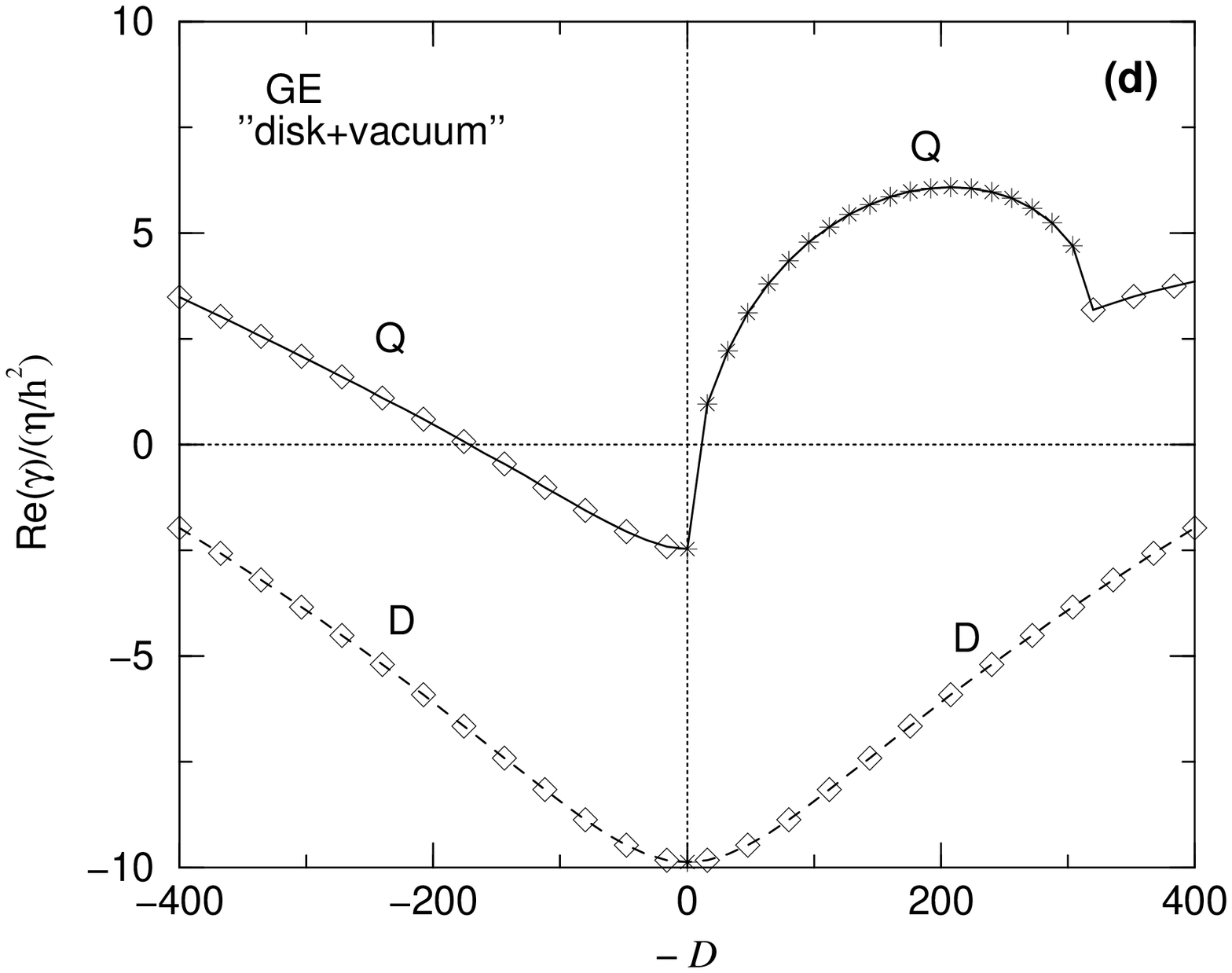}
\caption{
{\bf a-d} Real part of the growth rate of the magnetic field as a function of $- {\cal D}$
for $R_v = 0$.
Asterisks denote non-oscillatory solutions and diamonds denote oscillatory solutions.
Solid lines indicate quadrupolar modes and dashed lines indicate dipolar modes.
{\bf a} and {\bf c} are for the finite-difference model (FD) and {\bf b} and {\bf d} are 
for the
free-decay mode expansion model (GE) with {\bf a} and {\bf b} for the ``disk+halo'' model
and {\bf c} and {\bf d} for the ``disk+vacuum'' model.
Note the approximate symmetry of graphs {\bf a} and {\bf b} with respect to the vertical axis.
Note also that the dipolar mode in {\bf c} is the ``forgotten mode'', not seen in {\bf d}.}
\label{fig1}
\end{figure*}

We now want to compare the finite-difference simulation with the free-decay mode expansion. 
In particular,
we want to test our method to mimic, in the case of the free-decay mode expansion, 
a model
for $\eta_{\mathrm{halo}} = \eta_{\mathrm{disk}}$ starting with the ``disk+vacuum''
model (see Sect.~2). For an infinitely extended halo with 
$\eta_{\mathrm{halo}} = \eta_{\mathrm{disk}}$,
the growth rates of dipolar and quadrupolar modes should be exchanged when $\cal D$
reverses sign, i.e. the graph of $Re(\gamma)$ as a function of $\cal D$ should be
symmetric with respect to the vertical axis \citep{Pro77}. The finite-difference simulation 
shows this symmetry with a good precision (Fig.~\ref{fig1}a). The results of 
the free-decay mode expansion 
(Fig.~\ref{fig1}b) are less symmetric. This is not surprising since the ratio
between the height of the computational domain and the height of the disk is about
7 in the finite-difference simulation whereas it is only 2 in the free-decay 
mode expansion.

We now discuss our results for the ``disk+halo'' and ``disk+vacuum'' configuration. 
Let us first consider
the ``disk+halo'' configuration (Figs. \ref{fig1}a and b). For moderate dynamo number
$|{\cal D}| \lesssim 300$, the leading mode is quadrupolar and non-oscillatory
if ${\cal D} < 0$ ($R_\alpha > 0$) and it is dipolar and non-oscillatory if 
${\cal D} > 0$ ($R_\alpha < 0$). In the following, we thus study the effect
of vertical motions for ${\cal D} < 0$ in quadrupolar symmetry and for 
${\cal D} > 0$ in dipolar symmetry. Above $|{\cal D}| \simeq 300$, the symmetry
of the leading mode changes and it becomes oscillatory. The critical dynamo number is
$|{\cal D}_{\rm crit}| \simeq 5$.  

Let us now consider the ``disk+vacuum'' configuration (Figs.~\ref{fig1}c and d).
The results obtained with the two approaches are rather different. With the
free-decay mode expansion, the dipolar mode is always oscillatory and subcritical
for $|{\cal D}| \leq 400$. On the other hand, with the finite-difference simulation,
the dipolar mode becomes supercritical for ${\cal D} \gtrsim 50$ and is non-oscillatory
for all ${\cal D} > 0$. This is because the dominant dipolar mode in the
finite-difference simulation is the ``forgotten mode'' i.e. a dipolar mode 
corresponding to $n=0$ in Eqs.~(\ref{eqdipol1}) and (\ref{eqdipol2}). The 
``forgotten mode'' cannot be obtained with our 1D calculation (free-decay
mode expansion). 
However, we reproduced the corresponding profile $\gamma({\cal D})$
with a modified 1D model which takes into account the finite aspect
ratio of the disk. The model is based on the assumption of
separability of $r$- and $z$-dependence and yields equations
similar to the thin-disk equations (\ref{eq2}), (\ref{eq3}), but with additional
correction terms of order $(h/r)^2$.
Thus, the dipolar modes of Fig.~\ref{fig1}c and Fig.~\ref{fig1}d are
of a different nature. As for the quadrupolar mode, the results obtained with the
two approaches are not so different. The branch
for ${\cal D} < 0$ is rather similar in both cases with a critical dynamo
number $\gtrsim -50$. The branch for ${\cal D} > 0$  is all subcritical with 
the finite-difference
simulation whereas it becomes supercritical for ${\cal D} \sim 170$ with the free-decay
mode expansion. However, we have checked that, in the finite-difference model, if
we take a steeper transition between $\eta_{\mathrm{disk}}$ and 
$\eta_{\mathrm{halo}}$, the branch for ${\cal D} > 0$ moves upwards, resembling
the one for the free-decay mode expansion. 
 
The qualitative behaviour for the case of low conductivity in the halo is 
in many respects
similar to that of the ``disk+halo'' configuration (Figs.~\ref{fig1}a and c). 
But there are quantitative differences.
There is no intersection point where the symmetry of the leading mode would change
for $|{\cal D}| \lesssim 700$. The diagram does not look symmetric anymore.

The results of the finite-difference model presented here are in
agreement with the accretion disk models of \citet{RES95} and
\citet{vRE00}. Note, however, that these models have an inner
radius and that at this radius a perfectly conducting boundary condition
is used, which leads to a vertical current layer on the boundary and
potentially lowers the excitation threshold for magnetic fields.

\section{Linear behaviour with vertical velocities}
In this section, we study the linear behaviour of the $\alpha\omega$-dynamo 
in the presence of vertical velocity ($R_v \ne 0$). As discussed in Sect.~3,
we present results for quadrupolar symmetry with ${\cal D}<0$ (Sect.~4.1) and
results for dipolar symmetry with ${\cal D}>0$ (Sect.~4.2.). An explanation
of these results is given in Sect.~4.3.
 
\subsection{Quadrupolar modes}
\begin{figure*}
\centering
\includegraphics[angle=0,height=7.5cm,width=8.5cm]{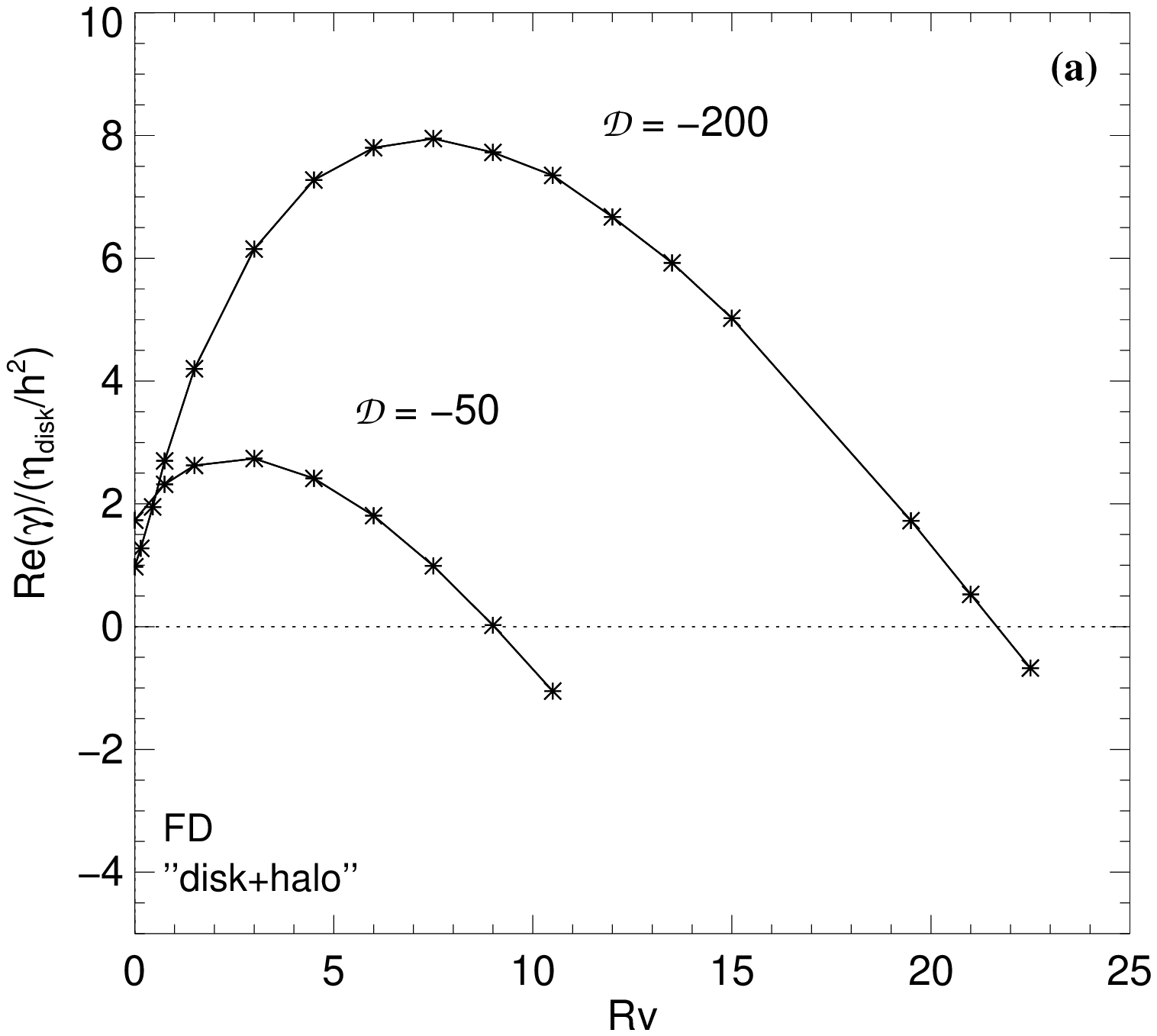}
\includegraphics[height=7.5cm,width=8.5cm]{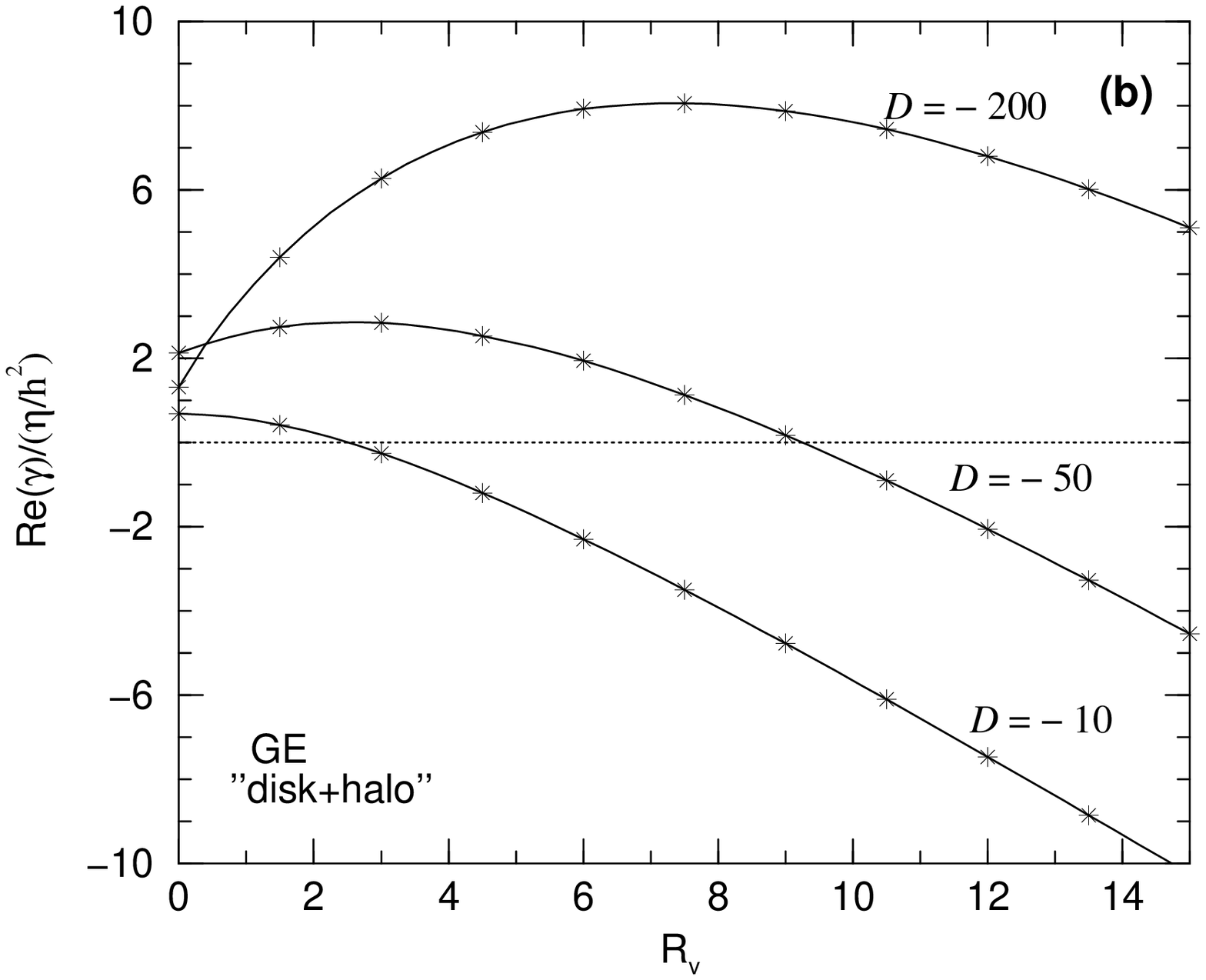}
\includegraphics[angle=0,height=7.5cm,width=8.5cm]{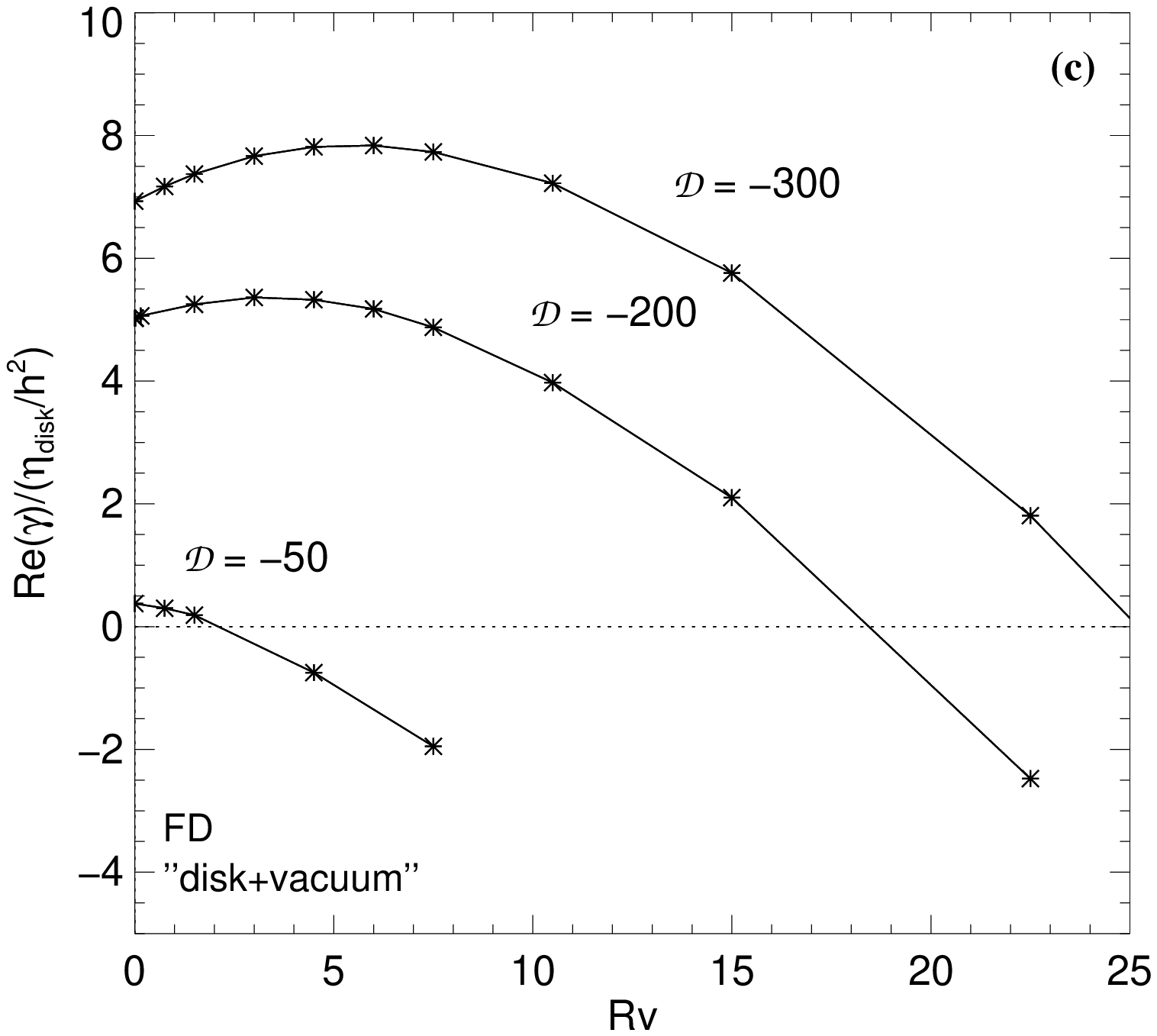}
\includegraphics[height=7.5cm,width=8.5cm]{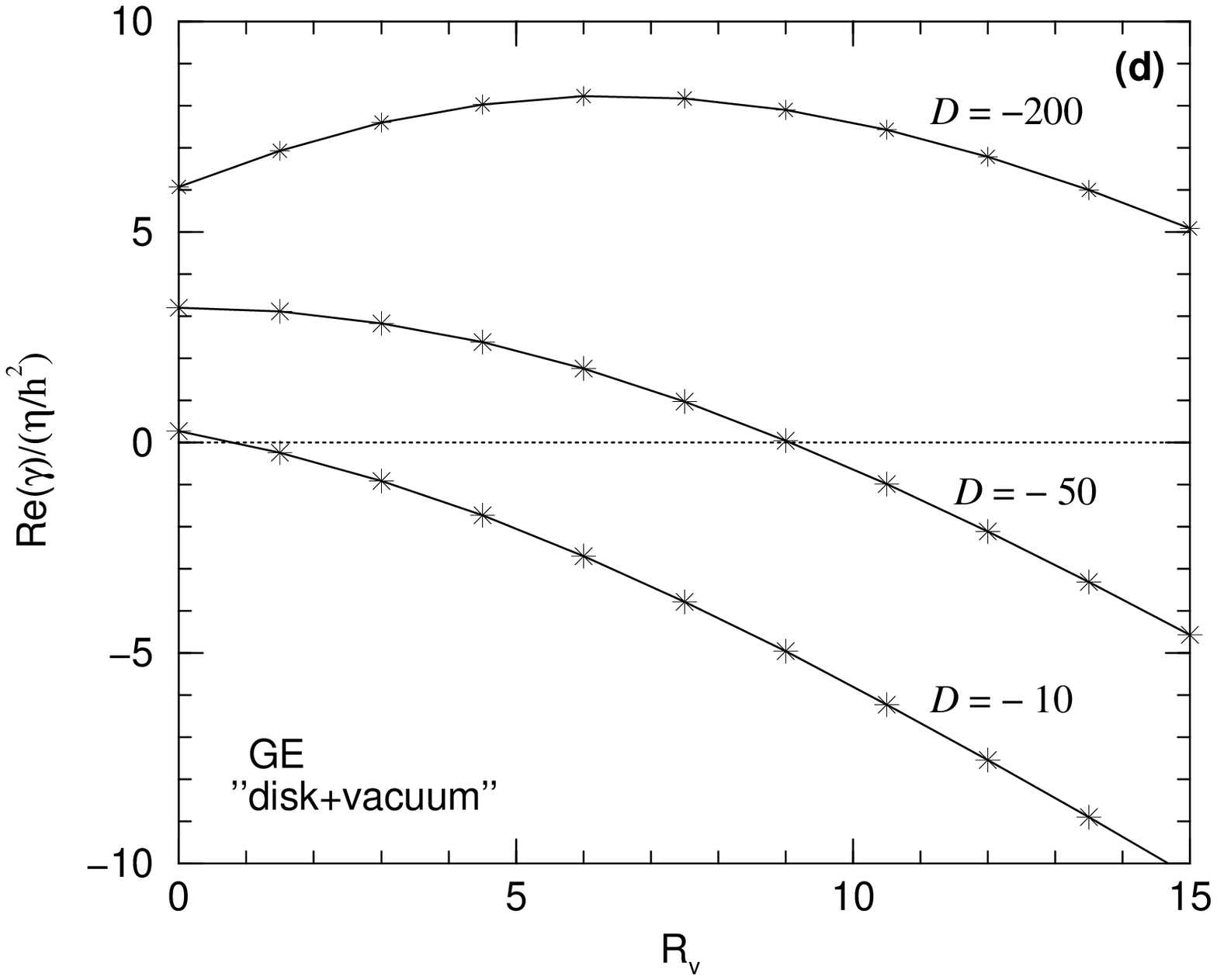}
\caption{{\bf a-d} Real part of the growth rate of the magnetic field as a function of $R_v$
for the dominant quadrupolar mode and for different dynamo numbers. All modes are 
non-oscillatory.
{\bf a} and {\bf c} are for the finite-difference model (FD) and {\bf b} and {\bf d} are for
the free-decay mode expansion model (GE) with {\bf a} and {\bf b} for the ``disk+halo''
model and {\bf c} and {\bf d} for the ``disk+vacuum'' model. The models of panels {\bf a} to
{\bf d} correspond to the ones in Fig.~\ref{fig1}.}
\label{fig2}
\end{figure*}

We consider quadrupolar modes with ${\cal D} < 0$.  
For sufficiently low dynamo numbers ($|{\cal D}| \lesssim 50$), the growth rate
decreases with $R_v$ whereas at larger dynamo numbers, the growth rate increases for
small $R_v$ and then decreases  at large $R_v$, forming a maximum (Fig.~\ref{fig2}). 
This maximum is a robust feature, the existence  of which does not seem
to depend on the boundary conditions or on the model. Indeed, we find this maximum with the
finite-difference simulation (Figs.~\ref{fig2}a and c) and with the free-decay mode expansion 
(Figs.~\ref{fig2}b and d) in both configurations, ``disk+halo''
and ``disk+vacuum'' (see Sect.~2.3 for the definition of these configurations).
The boundary conditions at the surfaces of the disk affect the position
of the maximum in $R_v$ and the dynamo number above which it occurs. In particular,
the smaller the magnetic diffusivity of the halo, the smaller this dynamo
number. Nevertheless, the required values for $\cal D$ to get a maximum are
larger than the typical values in galaxies. 
When ${\cal D}$ is increased, the maximum is more pronounced and its position
is shifted towards larger $R_v$. Note that the values of $R_v$ at which the
maximum occurs are between 1 and 10. They thus fit within the range
of values estimated for magnetic buoyancy.
Finally, in the range of dynamo numbers studied here, $|{\cal D}| \lesssim 300$, the
vertical velocity does not change the symmetry or temporal 
behaviour of the dominant mode.

\begin{figure*}
\centering
\includegraphics[height=7.5cm,width=8.5cm]{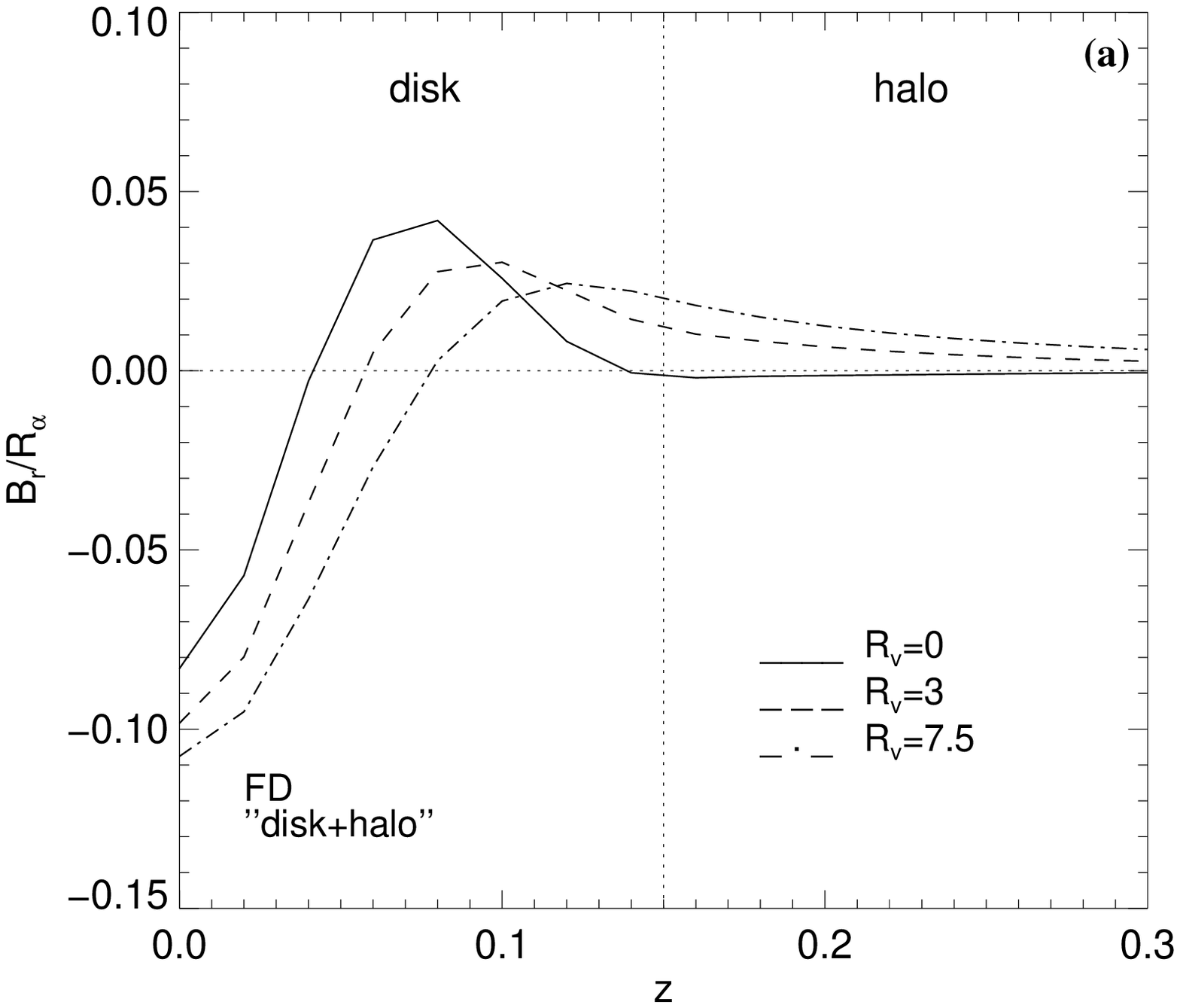}
\includegraphics[height=7.5cm,width=8.5cm]{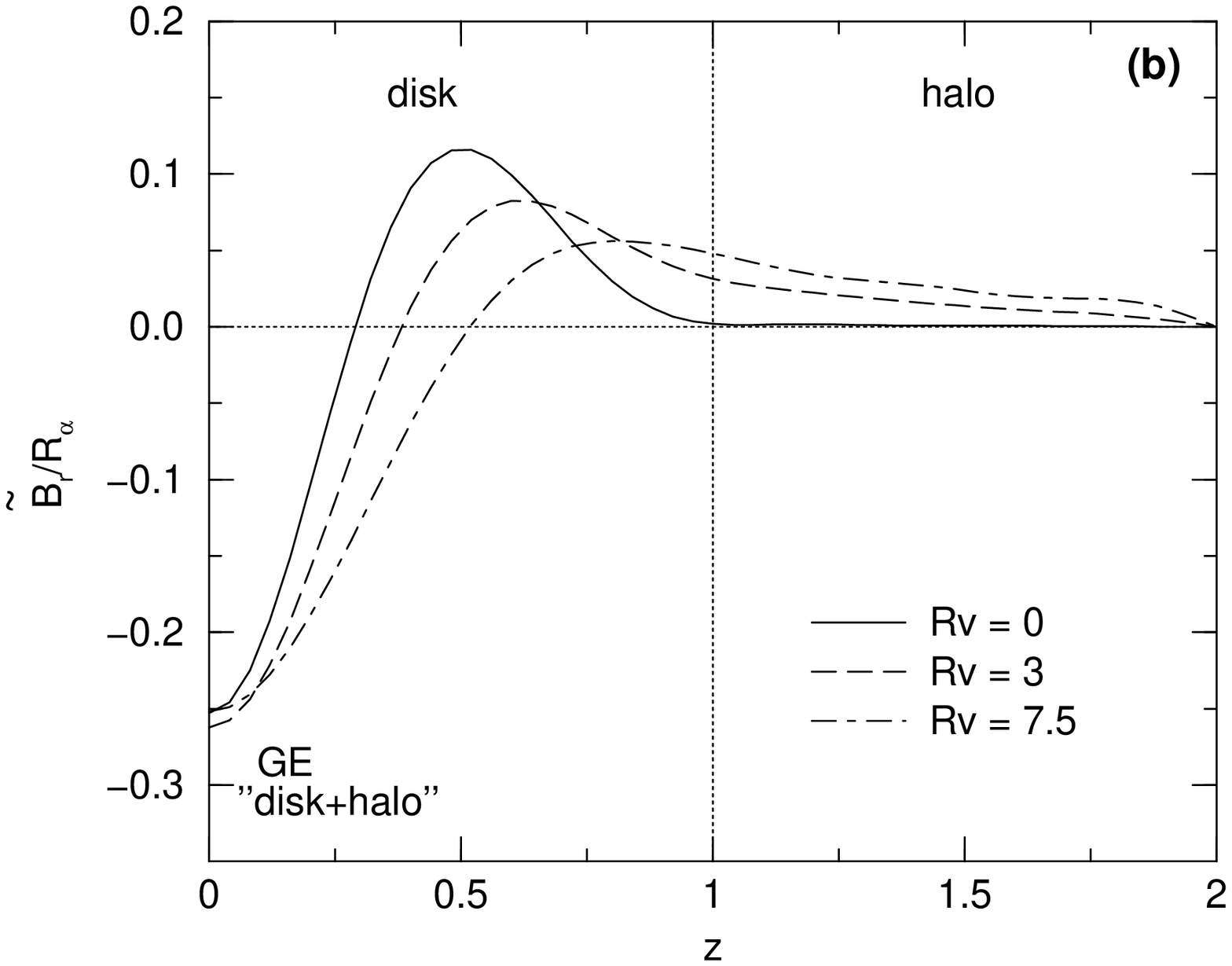}
\includegraphics[height=7.5cm,width=8.5cm]{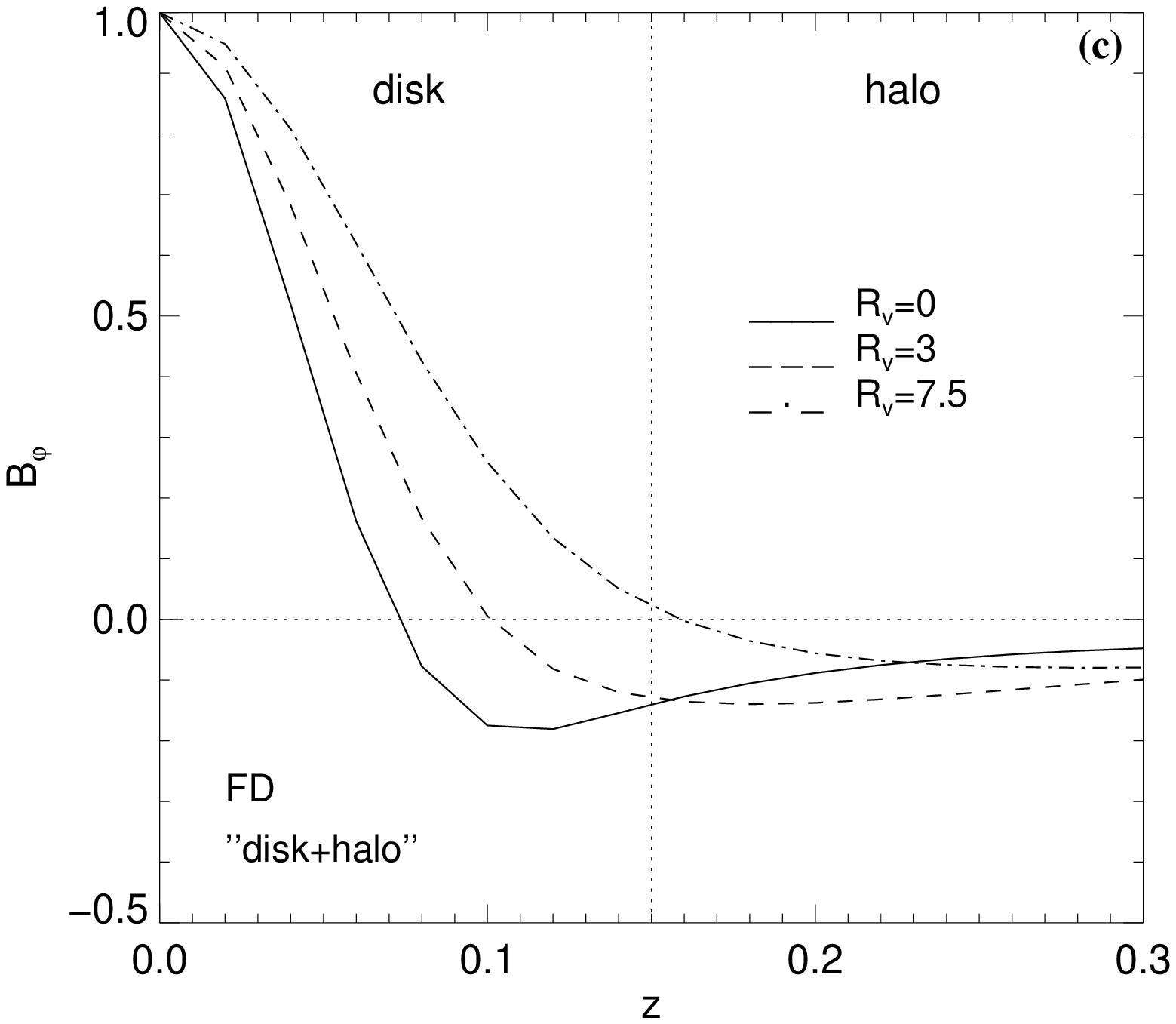}
\includegraphics[height=7.5cm,width=8.5cm]{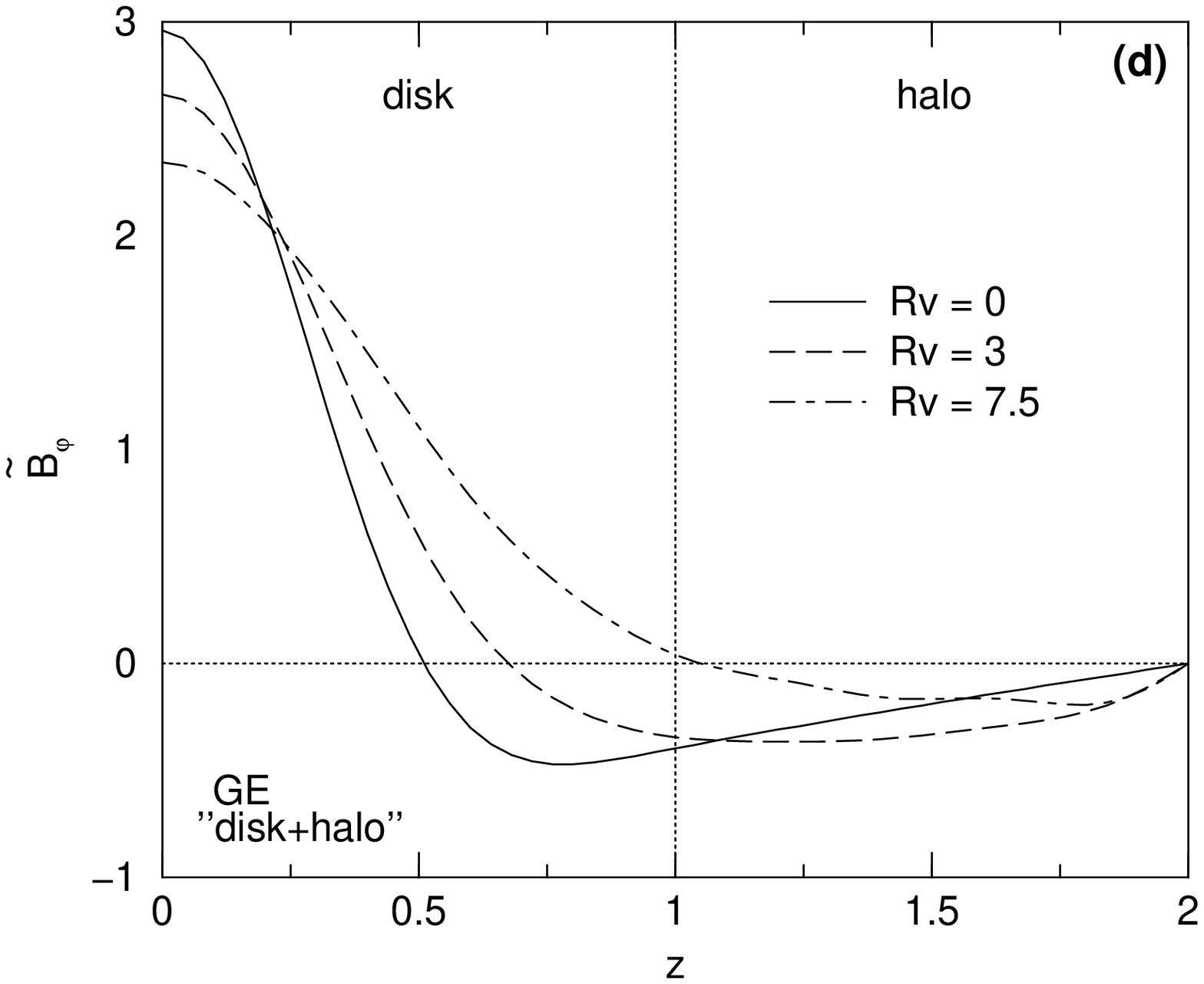}
\caption{{\bf a-d} Vertical structure of the non-oscillatory quadrupolar magnetic field
for ${\cal D} = -200$. Panels {\bf a} and {\bf c} are for the finite-difference model (FD)
while panels {\bf b} and {\bf d} are for the free-decay mode expansion (GE). Panels {\bf a} and
{\bf b} give the radial structure and {\bf c} and {\bf d} give the azimuthal structure of
the magnetic field. In the finite-difference model, eigenfunctions are taken 
at $r=1$ and are normalized such that $\max |B_\varphi| = 1$ for each $R_v$. Note that in 
the free-decay
mode expansion, we plot $\widetilde{B_r}$ and $\widetilde{B_{\varphi}}$, i.e. the
time-independent part of the magnetic field.}
\label{fig3}
\end{figure*}

We now discuss the magnetic field structure.
Figure~\ref{fig3} shows the eigenfunctions for $B_r$ and $B_{\varphi}$
for ${\cal D} = - 200$ and 
for different $R_v$ in the ``disk+halo'' configuration. 
With this choice of the dynamo
number, a maximum in the curve $\gamma(R_v)$ occurs around $R_v = 7.5$. An obvious 
feature is that the vertical
velocity pushes the zeros of $B_r$ and $B_{\varphi}$ outwards. For the dynamo to work,
it is essential that $B_r$ changes sign in the disk since only then can magnetic
flux of the opposite sign to that in the midplane leave through the disk surface
\citep{RSS88}. When $R_v$ takes large values, the radial component of the magnetic
field $B_r$ has no zero inside the disk and thus the dynamo cannot work ($\gamma < 0$).
This is the technical reason for the decrease of $\gamma$ with $R_v$. The physical
reason for it is investigated in Sect.~4.3. 
The increase of the growth rate at small $R_v$
will be explained in Sect.~4.3. We note however that when we decrease the truncation
level of the free-decay mode expansion down to $N=0$ (one doublet), the maximum disappears.

\begin{figure*}
\centering
\includegraphics[height=8.5cm,width=8.5cm]{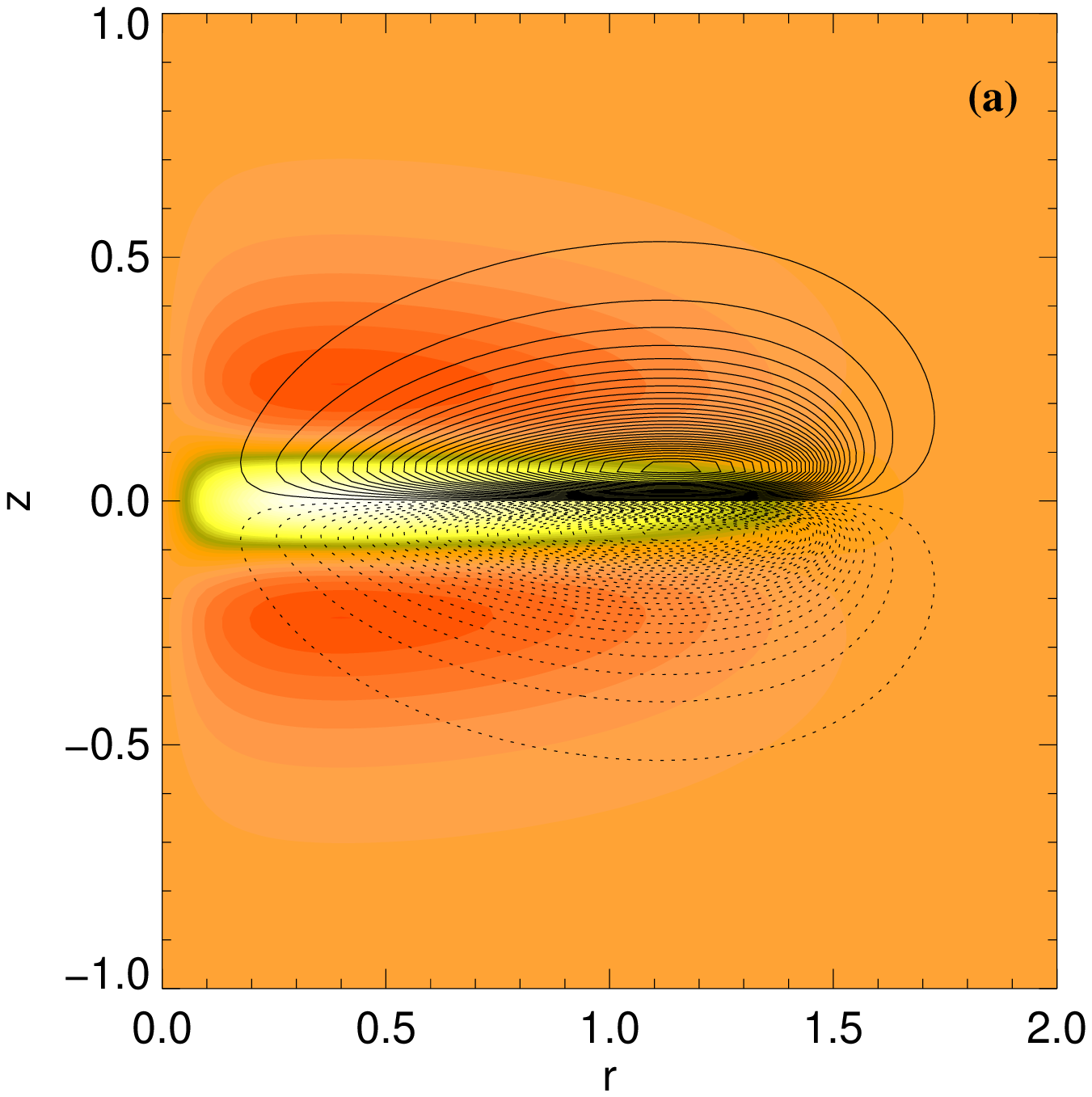}
\includegraphics[height=8.5cm,width=8.5cm]{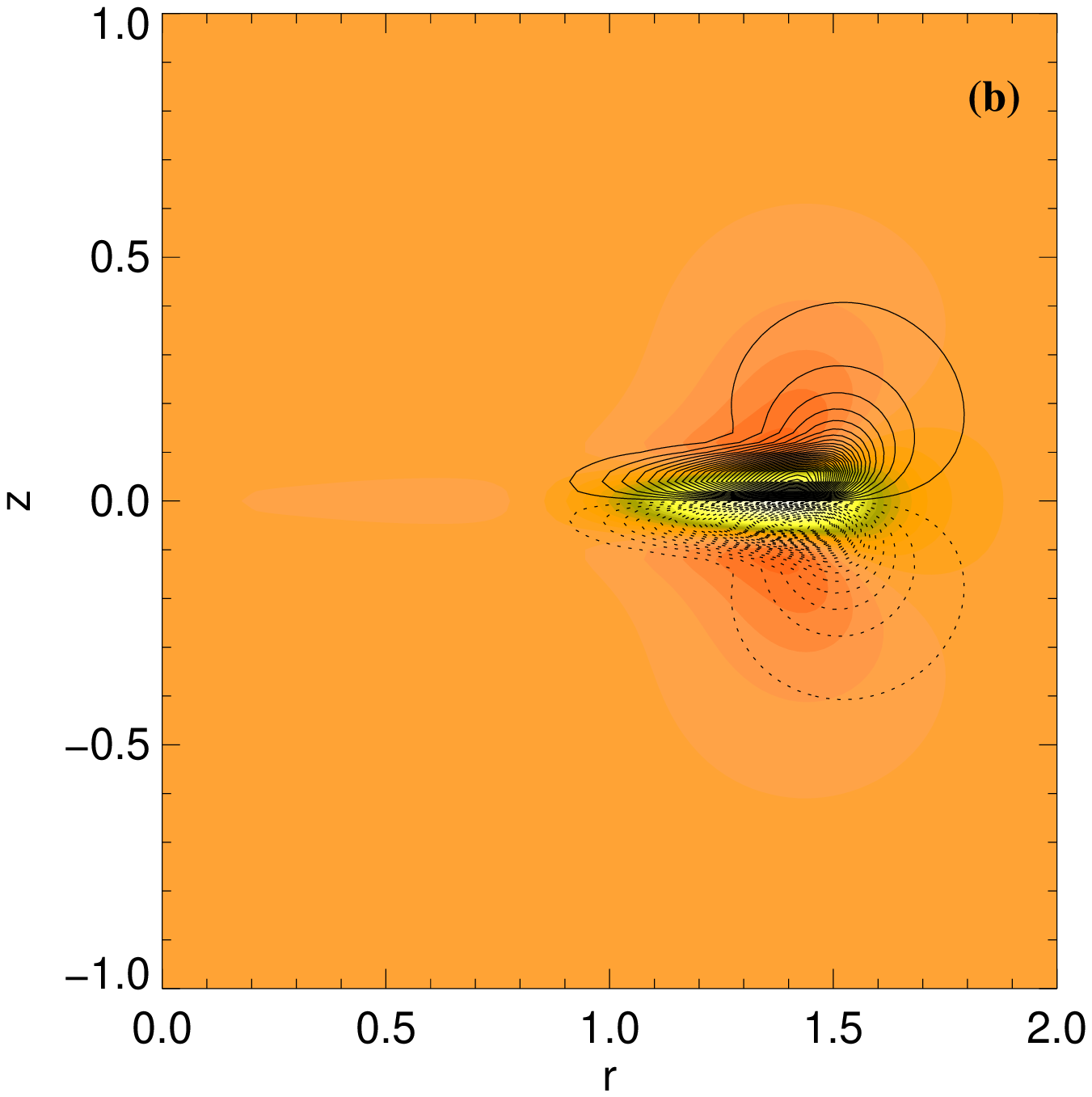}
\includegraphics[height=8.5cm,width=8.5cm]{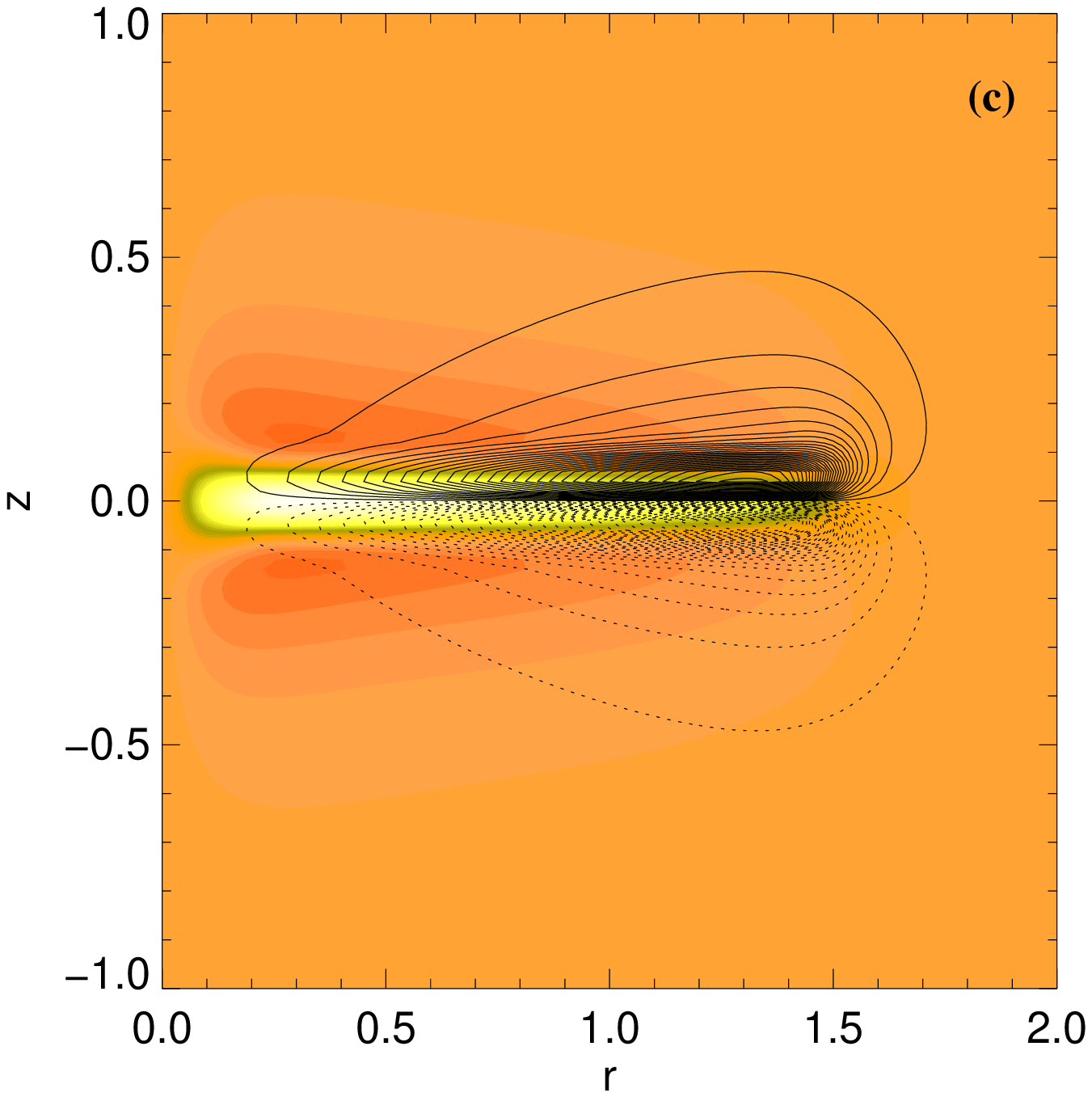}
\includegraphics[height=8.5cm,width=8.5cm]{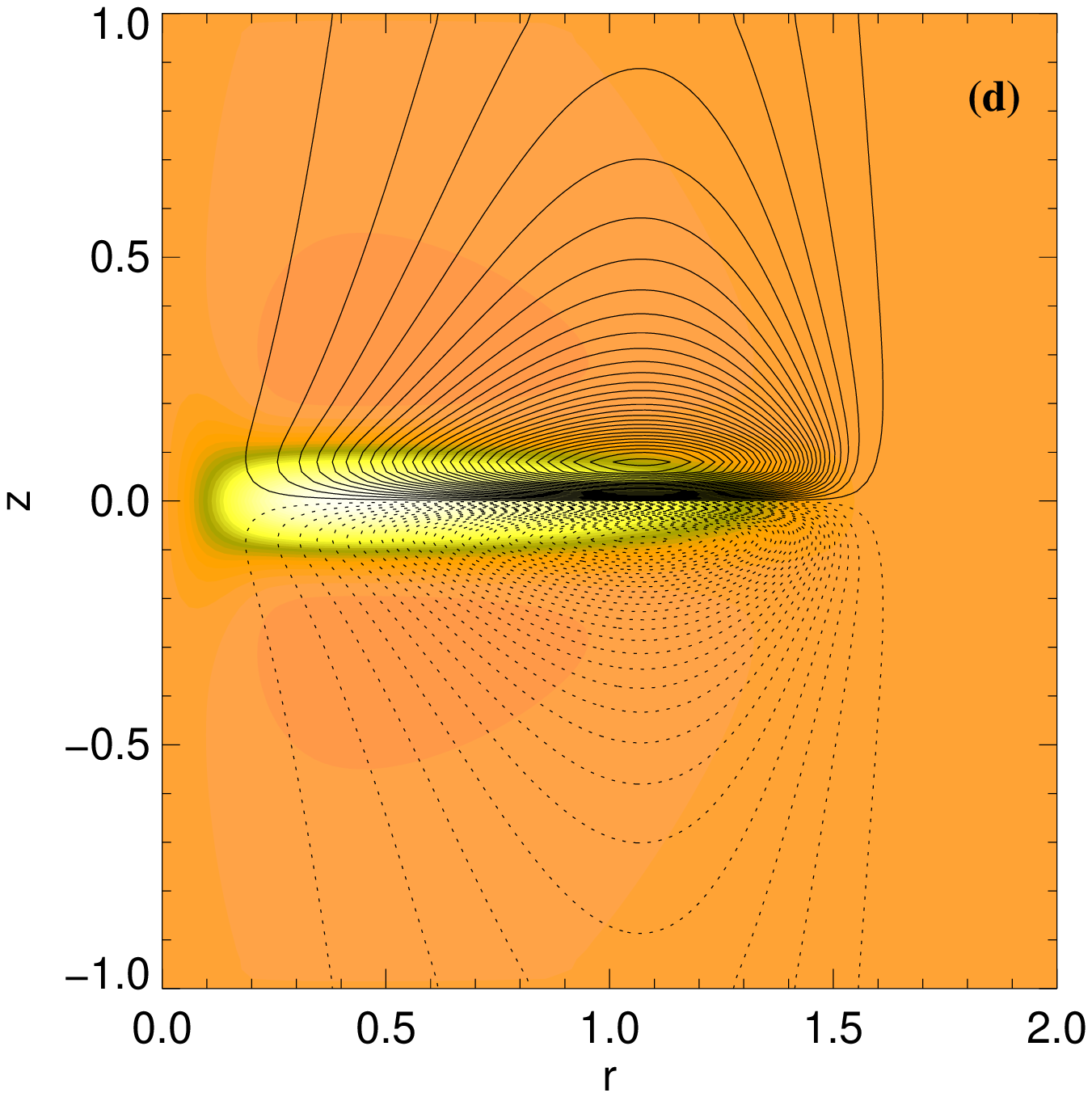}
\caption{{\bf a-d} Magnetic field configuration obtained using the finite-difference model
with the linear ``disk+halo'' configuration and for the dominant non-oscillatory
quadrupolar mode. 
Contours represent poloidal magnetic field lines with solid lines being clockwise
and dotted lines being counter-clockwise. The greyscales are the intensity and
direction of the azimuthal magnetic field; light grey (dark grey) signifies 
$B_\varphi > 0$ ($B_\varphi < 0$).
The parameters are {\bf a} ${\cal D} =  -50$, $R_v = 0$,
{\bf b} ${\cal D} = -200$, $R_v = 0$,
{\bf c} ${\cal D} = -200$, $R_v = 0.75$,
{\bf d} ${\cal D} = -200$, $R_v = 7.5$ (where Re$(\gamma)$ is maximum).
Comparing {\bf a} and {\bf b} shows the effect of increasing the dynamo number.
{\bf b}--{\bf d} show the effect of increasing $R_v$.}
\label{fig4}
\end{figure*}

The magnetic field configuration in 2D geometry is shown in
Fig.~\ref{fig4}. It shows both the effect of increasing the dynamo number 
and the effect of increasing the vertical magnetic Reynolds number. When one decreases
the dynamo number from ${\cal D} = -50$ to ${\cal D} = -200$, in the absence
of any vertical velocity, the magnetic field becomes more concentrated at
the outer edge of the disk (Figs.~\ref{fig4}a and \ref{fig4}b). 
If the vertical magnetic Reynolds number is then increased from $R_v = 0$,
the magnetic field is radially redistributed over the whole disk (Fig.~\ref{fig4}c)
and then advected towards the halo (Fig.~\ref{fig4}d).
From panel (b) to (d),
the growth rate increases as shown by the curve for ${\cal D} = -200$ in Fig.~\ref{fig2}a.
Thus, the magnetic field diffusion becomes less efficient as the vertical scale
of the magnetic field increases with $R_v$.
This leads to an increase of the growth rate up towards its maximum at $R_v = 7.5$
(Fig.~\ref{fig4}d).
Note that in the halo the poloidal magnetic field lines become more and more
aligned with lines of constant angular velocity as $R_v$ increases.
Note also that the mode structure of Fig.~\ref{fig4}d (where Re($\gamma$) is maximum
for ${\cal D} = -200$) is very similar to the mode structure for $- {\cal D} = 50, 
R_v = 3$, where that mode has attained its maximum growth rate.

\subsection{Dipolar modes}
\begin{figure}
\centering
\includegraphics[height=7cm,width=8cm]{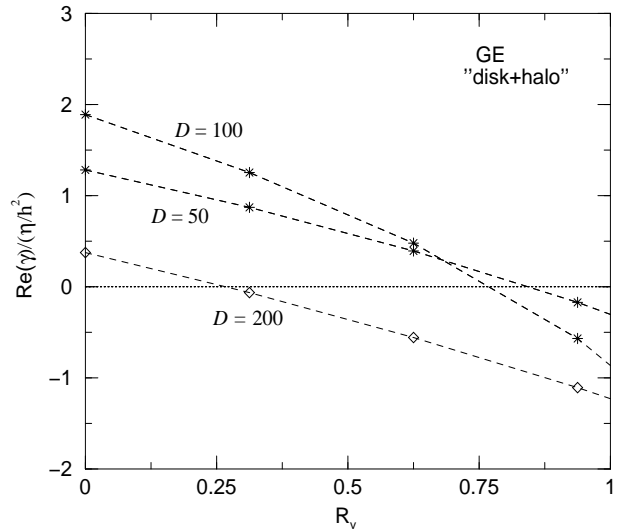}
\caption{Real part of the growth rate of the magnetic field as a 
function of $R_v$
for the dominant dipolar mode and for different dynamo numbers in the ``disk+halo''
model.
Asterisks denote non-oscillatory modes whereas diamonds denote oscillatory ones.}
\label{fig5}
\end{figure}

We now study the linear behaviour with vertical velocity for positive ${\cal D}$
where the dipolar non-oscillatory mode is dominant.
Figure~\ref{fig5} shows that there is no maximum in the growth rates; at $R_v$
of order $1$ or even less the dynamo is switched off.

\begin{figure}
\centering
\includegraphics[height=4.3cm,width=4.3cm]{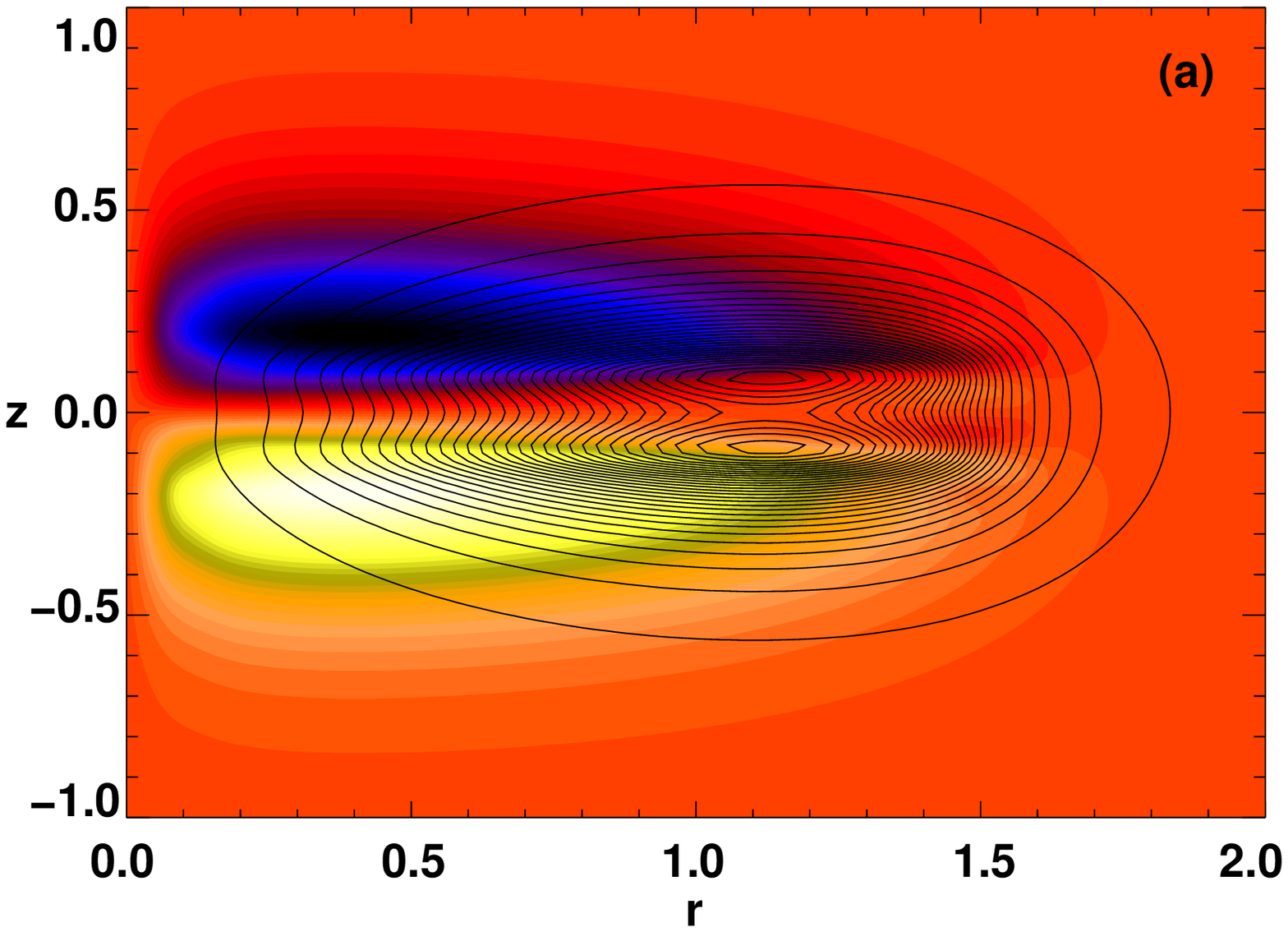}
\includegraphics[height=4.3cm,width=4.3cm]{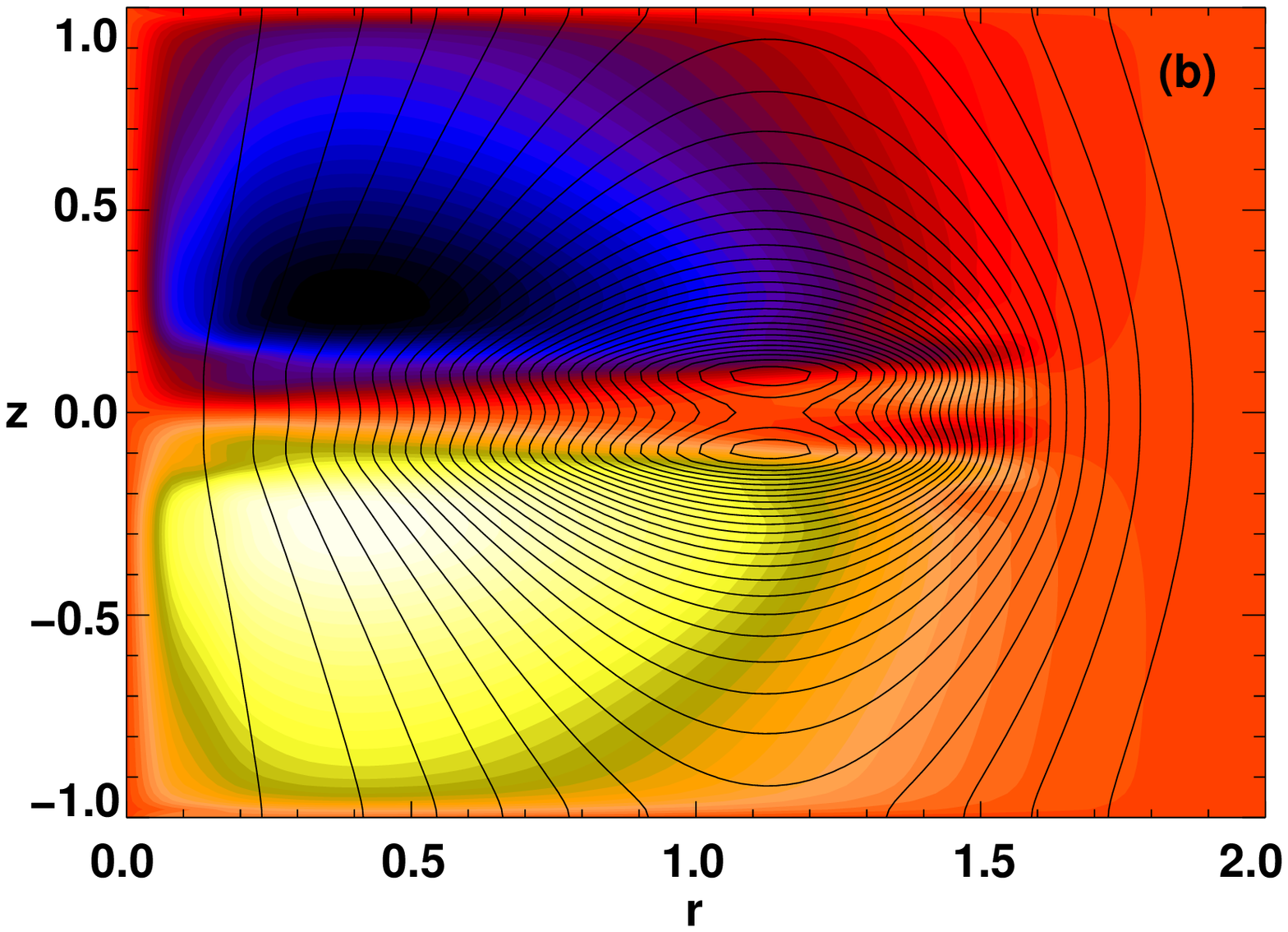}
\caption{{\bf a and b.} Magnetic field configuration obtained using the finite-difference model
with the linear ``disk+halo'' configuration and for the dominant non-oscillatory
dipolar mode. The parameters are  
{\bf a} ${\cal D} = 50$, $R_v = 0$,
{\bf b} ${\cal D} = 50$, $R_v = 0.75$. The notation is the same as in Fig.~\ref{fig4}.}
\label{fig4anum}
\end{figure}

The magnetic field structure in 2D geometry is shown in Fig~\ref{fig4anum}.
The magnetic field structures for ${\cal D} = 50$ and $R_v = 0$ and $0.75$ show
that the field is advected vertically outwards and the poloidal field lines are aligned 
with the $\Omega$ contours lines very quickly.

\subsection{The origin of the dynamo enhancement}
In this section, we explain why a vertical velocity can increase 
the growth rate of the dynamo. Therefore,
we plot the different terms of Eq.~(\ref{base1}) and Eq.~(\ref{base2}) obtained
with the free-decay mode expansion. To avoid any 
ambiguity with the choice of normalization, in particular when comparing
results at different $R_v$, we rewrite the equations in the following way.
We multiply Eq.~(\ref{base1}) and Eq.~(\ref{base2})  
by $B_r$ and $B_{\varphi}$ respectively, and we
integrate over the disk thickness. We thus obtain 
\begin{eqnarray}
\gamma = \gamma_\alpha^r + \gamma_{\mathrm{diff}}^r + \gamma_{\mathrm{stret}}^r,
\label{eqmax1}\\
\gamma = \gamma_\omega^\varphi + \gamma_{\mathrm{diff}}^\varphi
          + \gamma_{\mathrm{stret}}^\varphi,
\label{eqmax2}
\end{eqnarray}
with
\begin{eqnarray}
\gamma_\alpha^r &=& - R_\alpha \frac{\int_0^1 B_r (\alpha B_{\varphi})' \mathrm{d}z}
                  {\int_0^1 B_r^2 \mathrm{d}z},\\
\gamma_\omega^\varphi &=& R_\omega \frac{\int_0^1 B_r B_{\varphi} \mathrm{d}z}
                   {\int_0^1 B_{\varphi}^2 \mathrm{d}z},\\
\gamma_{\mathrm{diff}}^{r,\varphi} &=& \frac{\int_0^1 B_{r,\varphi} B_{r,\varphi}'' \mathrm{d}z}
                    {\int_0^1 B_{r,\varphi}^2 \mathrm{d}z} 
                     =- \frac{\int_0^1 \left(B_{r,\varphi}'\right)^2 \mathrm{d}z}
                      {\int_0^1 B_{r,\varphi}^2 \mathrm{d}z},\label{diss}\\ 
\gamma_{\mathrm{stret}}^{r,\varphi} &=& 
                     - R_v \frac{\int_0^1 B_{r,\varphi} (V_z B_{r,\varphi})' \mathrm{d}z}
                                   {\int_0^1 B_{r,\varphi}^2 \mathrm{d}z}
                         -\frac{R_v}{2} \frac{\int_0^1 V_z' B_{r,\varphi}^2 \mathrm{d}z}
                         {\int_0^1 B_{r,\varphi}^2 \mathrm{d}z}, 
\end{eqnarray}
where we have used a vacuum boundary condition at $z=h$ to rewrite 
$\gamma_{\mathrm{stret}}$, $\gamma_{\mathrm{diff}}$. This form of 
the equations allows us
to investigate the different physical contributions to the growth rate $\gamma$.
Note that Eqs.~(\ref{eqmax1}) and (\ref{eqmax2}) are not explicitly time-dependent. 
The terms $\gamma_\alpha^r$ and $\gamma_\omega^\varphi$ are the contributions to the
growth rate $\gamma$ from the $\alpha$-effect
and $\omega$-effect. The terms $\gamma_{\mathrm{diff}}^{r,\varphi}$ are 
the contributions from dissipation which are always negative. 
Finally, the terms $\gamma_{\mathrm{stret}}^{r,\varphi}$ are the contributions 
to $\gamma$ from 
stretching the magnetic field. The profile for $V_z$ is always antisymmetric
with respect to the disk midplane and $V_z' > 0$ for $z>0$. Thus, these
stretching terms $\gamma_{\mathrm{stret}}^{r,\varphi}$ are always negative. 
In particular, with a linear profile for
$V_z$, they are both equal to $-R_v/2$.

\begin{figure*}
\includegraphics[width=18cm]{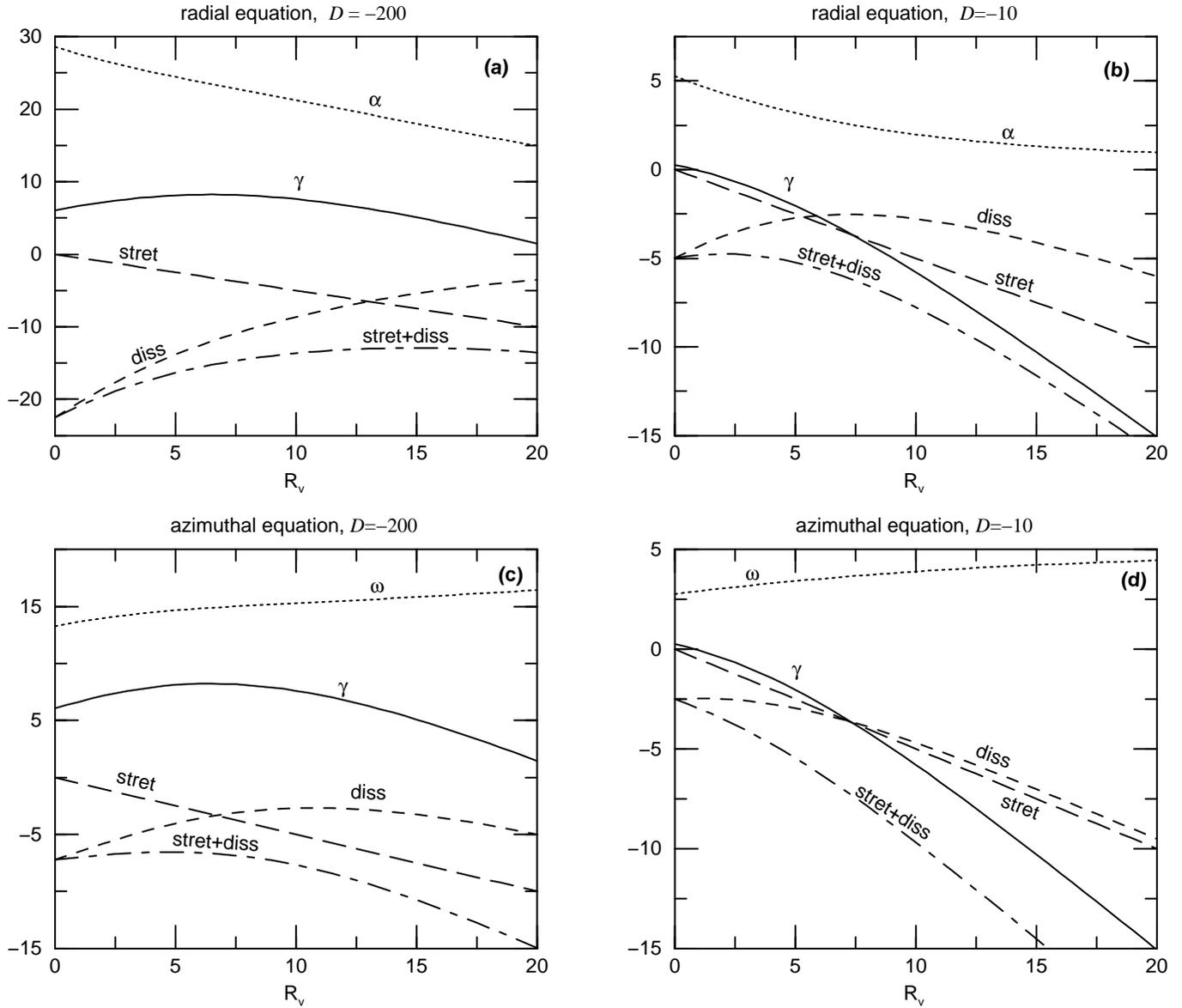}
\caption{{\bf a-d} Contributions of the different terms in Eqs.~(\ref{eqmax1}) 
and (\ref{eqmax2})
to the growth rate $\gamma$ in the ``disk+vacuum'' model. Lines are labelled
with the subscripts used for the different terms in Eqs.~(\ref{eqmax1}) and (\ref{eqmax2}).
{\bf a} and {\bf c} represent the radial and azimuthal terms for ${\cal D} = -200$, for
which a maximum occurs, {\bf b} and {\bf d} represent the radial and azimuthal terms for
${\cal D} = -10$, for which no maximum occurs.}
\label{figmax}
\end{figure*}

The general picture is as follows. The positive contribution to $\gamma$
comes from $\gamma_\alpha^r$ and $\gamma_\omega^\varphi$  whereas dissipation has
a negative contribution. In the presence of a vertical velocity,
an additional negative contribution comes from the stretching of the magnetic
field $\gamma_{\mathrm{stret}}^{r,\varphi}$. But, at the same time, the stretching 
reduces
dissipation by increasing the scale of the magnetic field (see Eq.~(\ref{diss})). 
Thus, depending on the relative values of the reduction of dissipation
and of the negative stretching term, the growth rate is enhanced or not (a
maximum occurs or not). Obviously, when the magnetic field is completely 
stretched, dissipation cannot be reduced anymore. However, this does not 
necessarily correspond to the maximum of the growth rate as the terms 
$\gamma_\alpha^r$ and $\gamma_\omega^\varphi$
evolve themselves with $R_v$ (which
quantifies the vertical velocity). These different quantities are plotted in
Fig.~\ref{figmax} in a case where a maximum occurs and in a case with no
maximum.

We note that dissipation increases again after being reduced by the stretching
of the magnetic field in the vertical direction. This increased dissipation
is due to the formation 
of boundary layers near $z=h$. We also note that the term $\gamma_\alpha^r$ 
decreases whereas $\gamma_\omega^\varphi$ increases.
The evolution of these terms is linked to the disappearance
of the positive area of $B_r$ (with the convention of sign of Fig.~\ref{fig3}). 
Finally, note that at large $R_v$, the growth rate $\gamma$ always decreases
as the stretching terms $\gamma_{\mathrm{stret}}^{r,\varphi}$ become dominant.

In short, the growth rate can increase with $R_v$, thus forming a maximum,
because dissipation is reduced by the stretching of the magnetic field.

\section{Nonlinear behaviour with vertical velocities}
The nonlinear behaviour has only been investigated with the finite-difference model.
We restrict ourselves here to the model ``disk+halo'' ($\eta_{\rm halo} = \eta_{\rm disk}$)
and to negative dynamo numbers considering only quadrupolar symmetry
which, for $R_v = 0$, is the dominant symmetry for moderate values of ${\cal D} < 0$.
Solutions are all saturated and non-oscillatory, as in the corresponding
linear modes.
We study three different cases of nonlinearity: magnetic buoyancy,
$\alpha$-quenching and a combination of both.
For a comparison of nonlinear solutions with a different parametrization of
magnetic buoyancy see earlier works by \citet{TBa94,TBb94}.

\subsection{Magnetic buoyancy}
We parameterize the effect of magnetic buoyancy by assuming that the vertical velocity
depends on the horizontal magnetic field,
\begin{equation}
V_z(z) = V_{z0} \, \frac{z}{h} \, \sqrt{B_r^2 + B_\varphi^2} / B_{\rm eq},
\label{eqnonlin-loc}
\end{equation}
where $B_{\rm eq}$ is the equipartition magnetic field 
$B_{\rm eq} = \sqrt{\rho \mu_0} c_\mathrm{s} {\rm Ma}_{\rm turb}$,
$\rho$ is the density, $c_\mathrm{s}$ the sound speed and ${\rm Ma}_{\rm turb}$ a turbulent
Mach number, treated here as a free parameter.
The profiles of density $\rho$ and sound speed $c_\mathrm{s}$ are taken from a
simple model of a cool accretion disk embedded in a hot corona 
(\citealp{BDS00}; see also \citealp{DBS99}).

\begin{figure}
\centering
\includegraphics[angle=0,width=8.5cm,keepaspectratio]{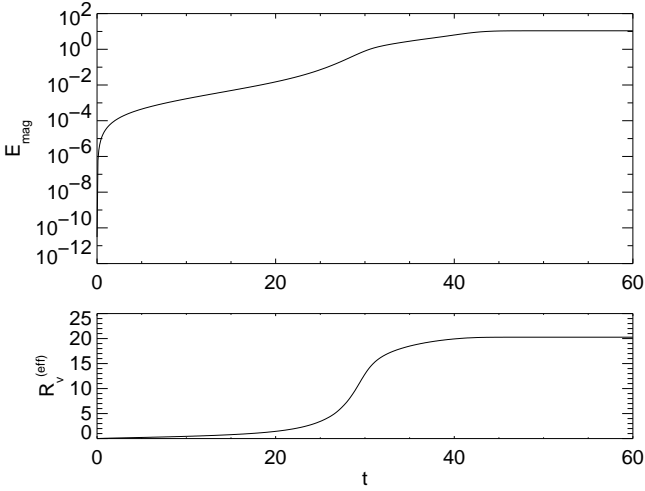}
\caption{Magnetic energy and effective vertical magnetic Reynolds number $R_v^{\rm (eff)}$
as a function of time for the ``disk+halo'' model with magnetic buoyancy 
(cf. Eq.~(\ref{eqnonlin-loc})).
${\cal D} = -200$ and $R_v = 7.5$.
Super-exponential growth occurs before saturation sets in.
The mode is quadrupolar and non-oscillatory.
The initial magnetic field  strength is $B_{\rm init}=10^{-5}$.
The steep increase for $t \lesssim 2$ is due to the generation of azimuthal field
from the initial large scale poloidal one at the chosen high $R_\omega$.}
\label{fig1721}
\end{figure}

We define an effective vertical magnetic Reynolds number
\begin{equation}
R_v^{\rm (eff)} = R_v \, \left\langle \sqrt{B_r^2 + B_\varphi^2} / B_{\rm eq} 
\right\rangle_{\rm disk},
\label{eqReff}
\end{equation}
where $\langle \cdot \rangle_{\mathrm{disk}}$ denotes averaging over the disk.
Fig.~\ref{fig1721} shows the magnetic energy and $R_v^{\rm (eff)}$ 
as a function of time for ${\cal D} = -200$ and $R_v = 7.5$.
This value of $R_v$ corresponds to the location of the maximum of the
growth rate $\gamma(R_v)$ in the linear model (cf. Fig.~\ref{fig2}a, upper curve).
Starting from an initial magnetic field strength $B_{\rm init}=10^{-5}$,
$\vec{B}$ grows and therefore $R_v^{\rm (eff)}$ 
increases; so the growth rate roughly follows the upper curve in Fig.~\ref{fig2}a.
Thus, as long as $R_v^{\rm (eff)}$ is approximately less than the position of the maximum
of $\gamma$, i.e. the horizontal magnetic field is smaller than the equipartition field,
$\gamma$ increases with time, which results in the
super-exponential growth of the magnetic energy seen for $10 \lesssim t \lesssim 28$
(Fig.~\ref{fig1721}). After $t \approx 28$, when the horizontal field is comparable
with the equipartition field, $\gamma$ decreases and eventually the magnetic field
approaches its saturation level. This level is reached when
$R_v^{\rm (eff)} \simeq R_{v*} \approx 21.5$, where $R_{v*}$
is the zero of $\gamma(R_v)$ in Fig.~\ref{fig2}a, upper curve.

In this subsection we assume that the only nonlinearity is due to magnetic buoyancy
and thus appears in the vertical velocity term only. In particular, we do not consider
here any dependence of $\alpha$ or $\eta$ on $\vec{B}$.
In this case the dimensionless form of the induction equation in the steady state reads
\begin{equation}
\hat{\cal L} \vec{B} + \vec{\nabla} \times
\left[\left(R_v \, \sqrt{B_r^2 + B_\varphi^2} / B_{\rm eq}\right) \,
\vec{u} \times \vec{B}\right] = 0,
\label{eqstst3}
\end{equation}
where $\hat{\cal L}$ is a linear operator and $u$ is a vector function
of position, $\vec{u} = (0,0,z/h)$ in the case considered here.
Let $\vec{b}$ be the solution of Eq.~(\ref{eqstst3}) for $R_v = 1$. Then, for arbitrary
$R_v$, the solution is $\vec{b}/R_v$.
From this, it follows that
the saturation value of magnetic energy is
\begin{equation}
E_{\mathrm{mag}}^{\mathrm{sat}} = {1 \over R_v^2} E_{\mathrm{mag}}^{\mathrm{sat}}\Bigr|_{R_v=1}
\propto {1 \over R_v^2},
\end{equation}
and the structure of the saturated magnetic field is independent of $R_v$.
This is confirmed with our numerical model. 

The magnetic field is extended
into the whole halo and almost completely stretched.
Therefore, adding a wind term such that
\begin{equation}
V_z(z) = \left(V_{z0}^{\rm buoy} \, \sqrt{B_r^2 + B_\varphi^2} / B_{\rm eq}
        + V_{z0}^{\rm wind}\right) \frac{z}{h}
\end{equation}
and fixing $R_v^{\rm buoy} = R_v^{\rm (eff)}$ while increasing $R_v^{\rm wind} = R_v$,
cannot lead to enhancement of magnetic energy. The wind is carrying magnetic field
away and $E_{\rm mag}$ is decreasing.

\subsection{$\alpha$-quenching}
We now parameterize the back-reaction of the magnetic field on the turbulent
$\alpha$-effect by making $\alpha$ depend on $\vec{B}$, 
\begin{equation}
\alpha(r,z) = \alpha_0(r) \sin\left(\pi \frac{z}{h}\right)
              {\xi_\alpha(r) \over 1 + {B^2 / B_{\rm eq}^2}},
\label{eqloc}
\end{equation}
where $B_{\rm eq}$ is again the equipartition magnetic field.
\begin{figure}
\centering
\includegraphics[angle=0,height=7.5cm,width=8.5cm]{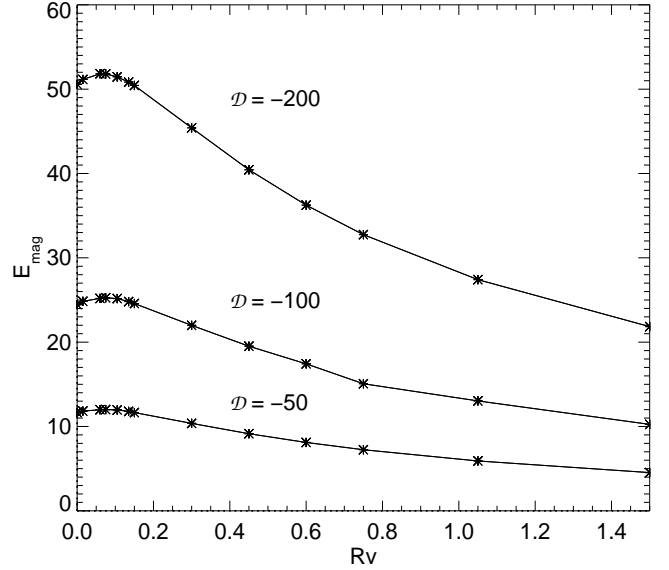}
\caption{Magnetic energy as a function of $R_v$ for the ``disk+halo'' model with
$\alpha$-quenching. All modes are quadrupolar and non-oscillatory.}
\label{fig9}
\end{figure}
\begin{figure*}
%\centering
\includegraphics[angle=0,height=5.9cm,width=5.9cm]{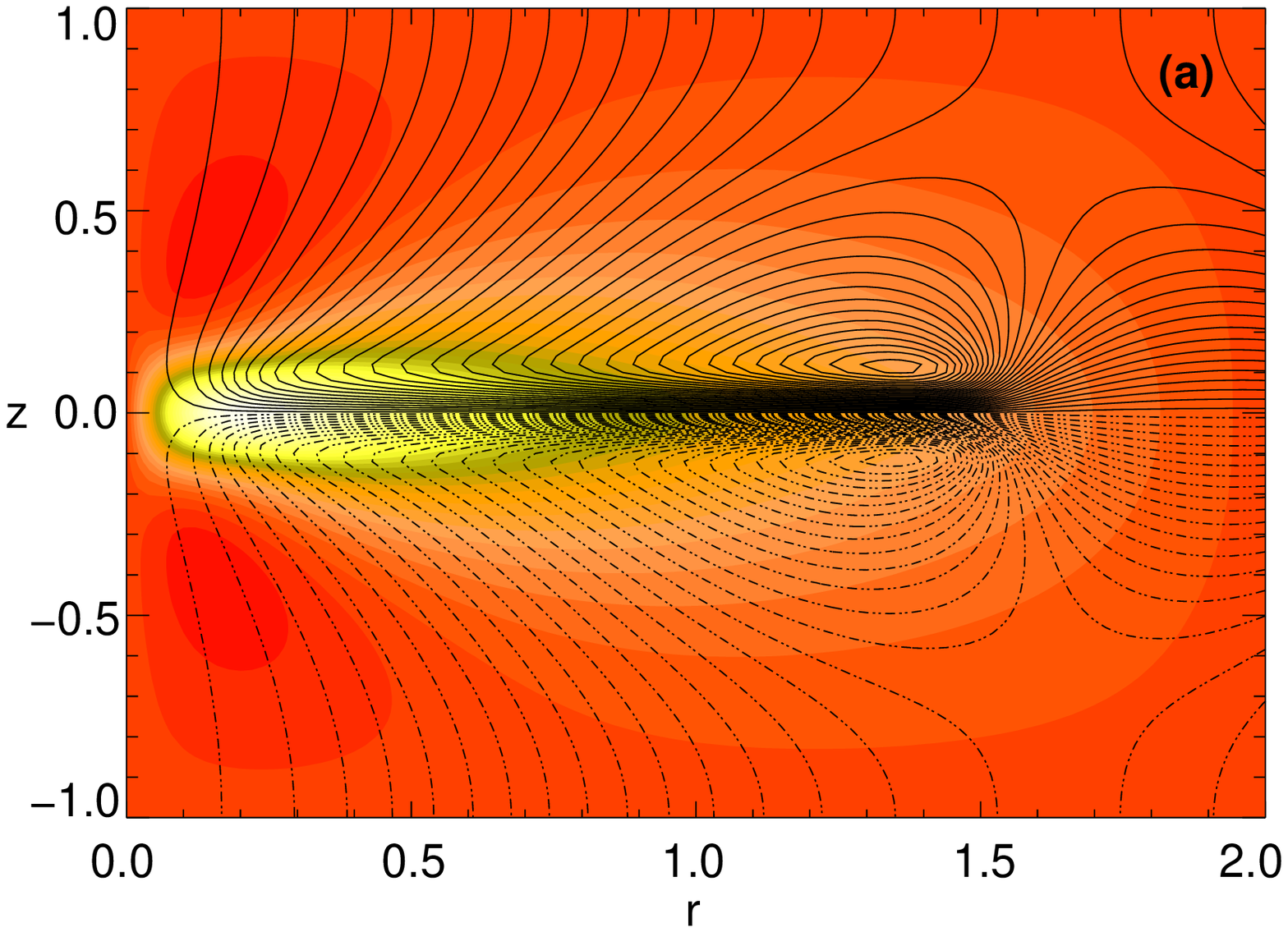}
\includegraphics[angle=0,height=5.9cm,width=5.9cm]{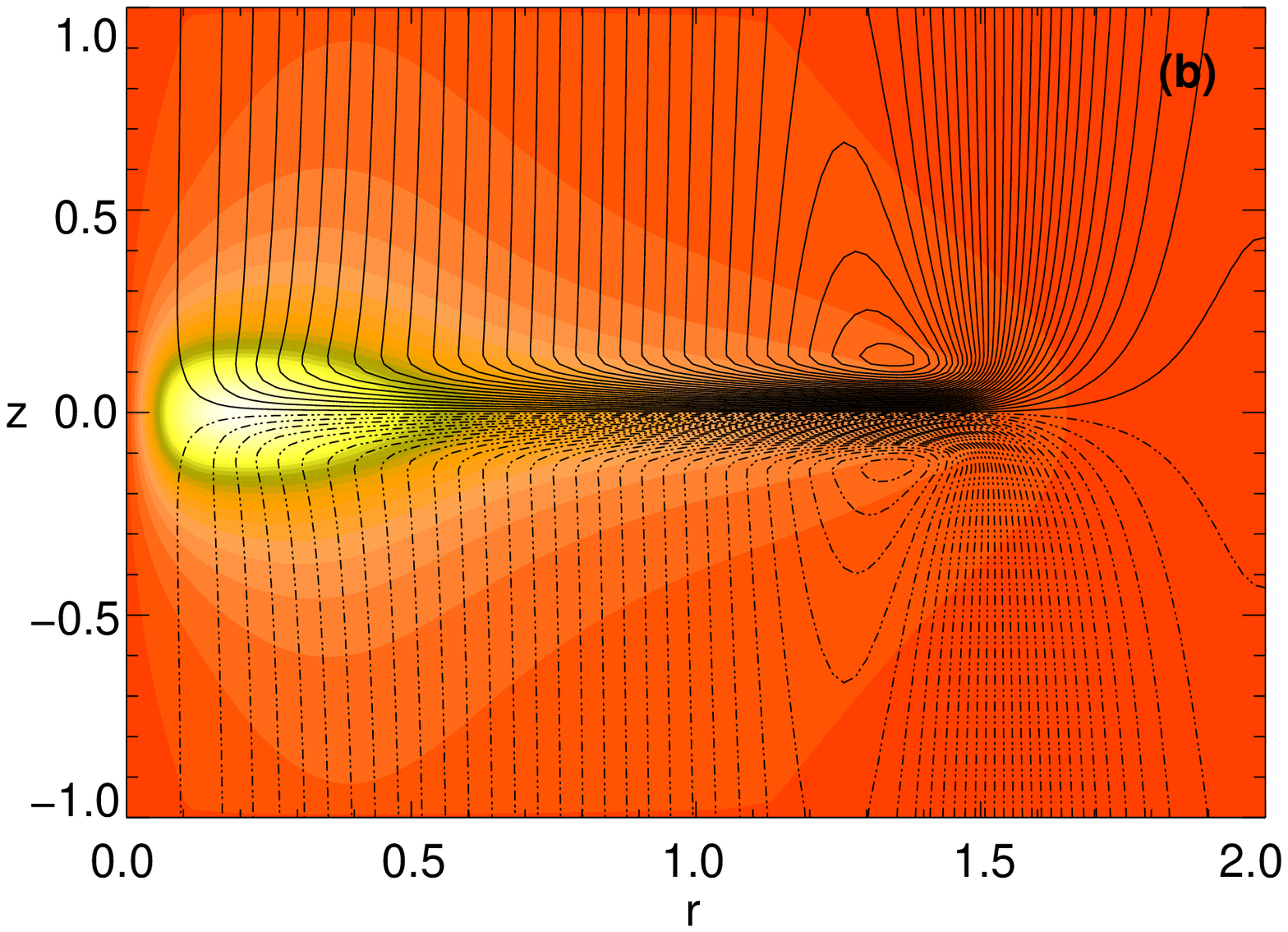}
\includegraphics[angle=0,height=5.9cm,width=5.9cm]{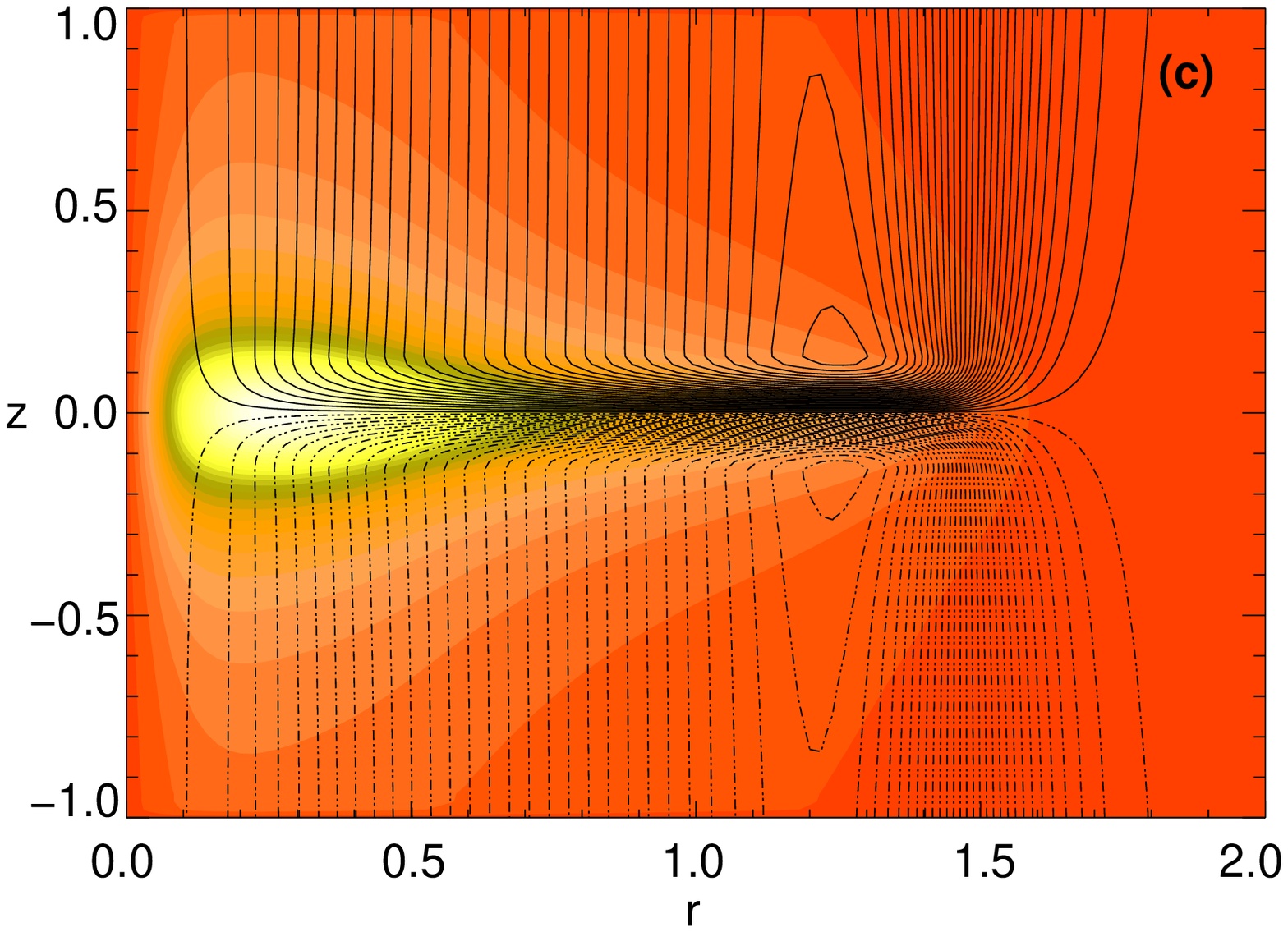}
\caption{{\bf a-c} Magnetic field configuration for the ``disk+halo'' model with 
$\alpha$-quenching: the effect of vertical velocity.
Poloidal field (lines) and azimuthal field (grey scales).
{\bf a} ${\cal D} = -50$, $R_v = 0$;
{\bf b} ${\cal D} = -50$, $R_v = 3$;
{\bf c} ${\cal D} = -50$, $R_v = 7.5$.
{\bf a-c} following the lower curve in Fig.~\ref{fig9}.
All modes are quadrupolar and non-oscillatory.}
\label{fig10}
\end{figure*}
Figure~\ref{fig9} shows the magnetic energy as a function of $R_v$ for the three
dynamo numbers ${\cal D} = -50$, ${\cal D} = -100$ and ${\cal D} = -200$.
From this we can see the following features:
\begin{itemize}
\item{$E_{\rm mag}$ has a maximum as a function of $R_v$ in the nonlinear regime;}
\item{but, this maximum is shifted towards much smaller $R_v$ (of order $0.1$), compared
to that of the growth rate in the linear regime, and it is also less pronounced;}
\item{the value of $R_v$ where the maximum in $E_{\rm mag}$ occurs, is almost independent of
${\cal D}$;}
\item{the magnetic energy scales roughly with the dynamo number,
$E_{\rm mag} \propto {\cal D}$;}
\item{all modes remain non-oscillatory when increasing $R_v$.}
\end{itemize}
The effect of vertical velocity on the structure of the magnetic field 
for the model with $\alpha$-quenching
can be seen in Fig.~\ref{fig10} where it is shown
for ${\cal D} = -50$. But looking at the magnetic field configuration
for ${\cal D} = -100$ and ${\cal D} = -200$, one finds that the field structure
changes only weakly with the dynamo number.

\subsection{$\alpha$-quenching together with magnetic buoyancy}
\begin{figure}
\centering
\includegraphics[angle=0,height=7.5cm,width=8.8cm]{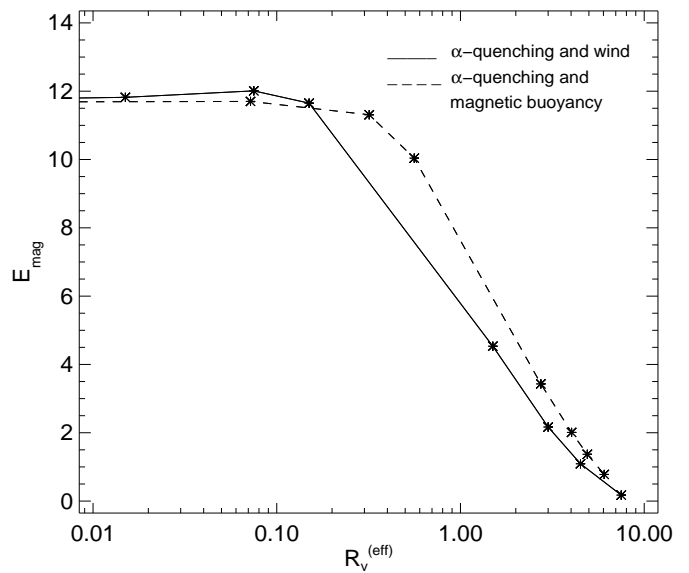}
\caption{Magnetic energy as a function of $R_v^{\rm (eff)}$ defined in
Eq.~(\ref{eqReff}) for the models
with $\alpha$-quenching with wind and $\alpha$-quenching combined with magnetic buoyancy.
$\eta_{\rm halo} = \eta_{\rm disk}$ and ${\cal D} = -50$.
All modes are quadrupolar and non-oscillatory.
With $\alpha$-quenching alone, $R_v^{\rm (eff)} = R_v$.}
\label{fig19}
\end{figure}

In this section, we combine $\alpha$-quenching, Eq.~(\ref{eqloc}), with magnetic
buoyancy, Eq. (\ref{eqnonlin-loc}), and compare the results with the case of
$\alpha$-quenching with wind.
In Fig.~\ref{fig19}, magnetic energy is plotted against $R_v^{({\rm eff})}$ for
  these two models and for ${\cal D} = -50$.
Evidently, the two curves behave similarly, supporting the notion that
wind and magnetic buoyancy have similar effects on the dynamo.
For $R_v^{({\rm eff})} \ll 1$, the field structure is solely determined by
$\alpha$-quenching and thus $E_{\rm mag}$ is independent of
$R_v^{({\rm eff})}$.
For $R_v^{({\rm eff})} \gg 1$, on the other hand, $\alpha$-quenching
becomes unimportant and $E_{\rm mag} \propto (R_v^{({\rm eff})})^{-2}$, as
discussed in Sect. 5.1.
The largest disagreement between the two curves occurs for intermediate
values of $R_v^{({\rm eff})}$, i.e. where both $\alpha$-quenching and
magnetic buoyancy are important, and where the field has not yet reached
its asymptotic shape (Fig.~\ref{fig10}c).

Perfect agreement between the two curves cannot be expected, because the
averaging in Eq.~(\ref{eqReff}) was somewhat arbitrary; the agreement could
perhaps be improved by choosing a different definition of $R_v^{({\rm eff})}$.
The fact that with magnetic buoyancy the magnetic energies are larger for
a given value of $R_v^{({\rm eff})}$ can therefore not be interpreted as a
relative enhancement of dynamo action.

\section{Conclusions}

Disk dynamos have been studied intensively over the past 20 years, but
only recently the effects of outflows and winds have been included
(e.g. \citealt{Cam99,DBS99,BDS00}).
It was immediately clear that these changes in the boundary conditions
affect the structure of the solutions.
The present work is an attempt to provide a more thorough survey of
solutions with wind for different values and sign of the dynamo number.
One of the effects that we have highlighted is the enhancement of
the kinematic dynamo growth rate for intermediate strengths of the wind.
In the nonlinear case this corresponds to a phase shortly before saturation
where the growth becomes super-exponential.

In the present work we have deliberately isolated the dynamo problem
from the wind problem. In reality the two aspects together
constitute a more complicated self-consistent dynamo problem that
needs to be looked at in more detail in future. One step in
that direction was taken by \citet{BDS00} where the
feedback on the flow from the Lorentz force was taken fully into account. A more
formidable task would be to obtain self-consistent dynamo action in
three dimensions without assuming an $\alpha$-effect, thus obtaining
dynamo action and outflows in a self-consistent manner from first
principles.
In the context of accretion disks, this would
require simulating turbulent dynamo action, where the turbulence itself
is driven by dynamo-generated magnetic fields \citep{BNS95}.
This would be of great interest not only from the astrophysical point
of view, but also from that of dynamo theory in general, because,
as was recently realized, the catastrophic quenching of the standard
$\alpha\omega$-dynamo could be alleviated in the presence of open
boundary conditions which may substantially improve the working of the 
dynamo \citep{BF00}.
This is simply because the $\alpha$-effect generates helical fields,
but at the same time magnetic helicity is conserved in a closed
domain. Thus, being able to get rid of helical fields through open
boundaries may be vital for the dynamo \citep{BD01}. Furthermore, the production
of fields at large scales could be facilitated by the production
-- and subsequent loss through boundaries -- of small scale fields
with opposite sign of magnetic helicity. This latter aspect could
not be addressed with mean-field models and would require a direct
simulation of accretion disk turbulence in a global model.

\acknowledgement{We acknowledge financial support from PPARC
 (Grant PPA/G/S/1997/00284) and from the Leverhulme Trust (Grant F/125/AL).
We thank D. Sokoloff for helpful discussions. The use of the PPARC
supported parallel computer GRAND in Leicester and the PPARC funded
Compaq MHD Cluster in St Andrews is acknowledged.}

\bibliographystyle{apj}
\bibliography{biblio}

\begin{thebibliography}{31}
\expandafter\ifx\csname natexlab\endcsname\relax\def\natexlab#1{#1}\fi

\bibitem[{Balbus \& Hawley(1998)}]{BH98}
Balbus, S.~A. \& Hawley, J.~F. 1998, Rev. Mod. Phys., 70, 1

\bibitem[{Blackman \& Field(2000)}]{BF00}
Blackman, E.~G. \& Field, G.~F. 2000, ApJ, 534, 984

\bibitem[{Brandenburg(1998)}]{Bra98}
Brandenburg, A. 1998, in Theory of black hole accretion discs, ed.
  M.~Abramowicz, G.~Bj{\"o}rnsson, \& J.~Pringle (Cambridge: Cambridge
  University Press), 61--90

\bibitem[{Brandenburg(2001)}]{Bra01}
Brandenburg, A. 2001, ApJ, in press (issue 550, 1 April), astro-ph/0006186

\bibitem[{Brandenburg \& Dobler(2001)}]{BD01}
Brandenburg, A. \& Dobler, W. 2001, A\&A, in press, astro-ph/0012472

\bibitem[{Brandenburg {et~al.}(2000)Brandenburg, Dobler, Shukurov, \& von
  Rekowski}]{BDS00}
Brandenburg, A., Dobler, W., Shukurov, A., \& von Rekowski, B. 2000, in
  preparation

\bibitem[{Brandenburg {et~al.}(1993)Brandenburg, Donner, Moss, Shukurov,
  Sokoloff, \& Tuominen}]{Bra93}
Brandenburg, A., Donner, K.~J., Moss, D., Shukurov, A., Sokoloff, D.~D., \&
  Tuominen, I. 1993, A\&A, 271, 36

\bibitem[{Brandenburg {et~al.}(1995)Brandenburg, Nordlund, Stein, \&
  Torkelsson}]{BNS95}
Brandenburg, A., Nordlund, {\AA}., Stein, R.~F., \& Torkelsson, U. 1995, ApJ,
  446, 741

\bibitem[{Campbell(1999)}]{Cam99}
Campbell, C.~G. 1999, MNRAS, 310, 1175

\bibitem[{Campbell {et~al.}(1998)Campbell, Papaloizou, \& Agapitou}]{CPA98}
Campbell, C.~G., Papaloizou, J. C.~B., \& Agapitou, V. 1998, MNRAS, 300, 315

\bibitem[{Dobler {et~al.}(1999)Dobler, Brandenburg, \& Shukurov}]{DBS99}
Dobler, W., Brandenburg, A., \& Shukurov, A. 1999, in Plasma turbulence and
  energetic particles in astrophysics, ed. M.~Ostrowski \& R.~Schlickeiser
  (Cracow: Obserwatorium Astronomiczne), 347--356

\bibitem[{Elstner {et~al.}(1995)Elstner, Golla, R{\"u}diger, \&
  Wielebinski}]{EGR95}
Elstner, D., Golla, G., R{\"u}diger, G., \& Wielebinski, R. 1995, A\&A, 297, 77

\bibitem[{Gabov {et~al.}(1996)Gabov, Sokoloff, \& Shukurov}]{GSS96}
Gabov, A.~S., Sokoloff, D.~D., \& Shukurov, A.~M. 1996, Astronomy Reports, 40,
  463

\bibitem[{Heyvaerts {et~al.}(1996)Heyvaerts, Priest, \& Bardou}]{HPB96}
Heyvaerts, J., Priest, E.~R., \& Bardou, A. 1996, ApJ, 473, 403

\bibitem[{Lubow {et~al.}(1994)Lubow, Papaloizou, \& Pringle}]{LPP94}
Lubow, S.~H., Papaloizou, J. C.~B., \& Pringle, J.~E. 1994, MNRAS, 267, 235

\bibitem[{Moss {et~al.}(1999)Moss, Shukurov, \& Sokoloff}]{MSS99}
Moss, D., Shukurov, A., \& Sokoloff, D. 1999, A\&A, 343, 120

\bibitem[{Moss {et~al.}(2000)Moss, Shukurov, \& Sokoloff}]{MSS00}
---. 2000, A\&A, 358, 1142

\bibitem[{Poezd {et~al.}(1993)Poezd, Shukurov, \& Sokoloff}]{PSS93}
Poezd, A., Shukurov, A., \& Sokoloff, D. 1993, MNRAS, 264, 285

\bibitem[{Pouquet {et~al.}(1976)Pouquet, Frisch, \& L\'eorat}]{PFL76}
Pouquet, A., Frisch, U., \& L\'eorat, J. 1976, J. Fluid. Mech., 77, 321

\bibitem[{Proctor(1977)}]{Pro77}
Proctor, M. R.~E. 1977, Astron. Nachr., 298, 19

\bibitem[{Pudritz(1981)}]{Pud81}
Pudritz, R.~E. 1981, MNRAS, 195, 897

\bibitem[{R{\"u}diger {et~al.}(1995)R{\"u}diger, Elstner, \& Stepinski}]{RES95}
R{\"u}diger, G., Elstner, D., \& Stepinski, T. 1995, A\&A, 298, 934

\bibitem[{Ruzmaikin {et~al.}(1988)Ruzmaikin, Shukurov, \& Sokoloff}]{RSS88}
Ruzmaikin, A.~A., Shukurov, A.~M., \& Sokoloff, D.~D. 1988, Magnetic fields of
  galaxies (Dordrecht: Kluwer Academic Publishers)

\bibitem[{Sokoloff \& Shukurov(1990)}]{SS90}
Sokoloff, D. \& Shukurov, A. 1990, Nature, 347, 51

\bibitem[{Stepinski \& Levy(1988)}]{SL88}
Stepinski, T.~F. \& Levy, E.~H. 1988, ApJ, 331, 416

\bibitem[{Stepinski \& Levy(1991)}]{SL91}
---. 1991, ApJ, 379, 343

\bibitem[{Torkelsson \& Brandenburg(1994{\natexlab{a}})}]{TBa94}
Torkelsson, U. \& Brandenburg, A. 1994{\natexlab{a}}, MNRAS, 283, 677

\bibitem[{Torkelsson \& Brandenburg(1994{\natexlab{b}})}]{TBb94}
---. 1994{\natexlab{b}}, MNRAS, 292, 341

\bibitem[{Vainshtein \& Cattaneo(1992)}]{VC92}
Vainshtein, S. \& Cattaneo, F. 1992, ApJ, 393, 165

\bibitem[{van Ballegooijen(1989)}]{vBa89}
van Ballegooijen, A.~A. 1989, in Accretion disks and magnetic fields in
  astrophysics, ed. G.~Belvedere (Dordrecht: Kluwer Academic Publishers),
  99--106

\bibitem[{von Rekowski {et~al.}(2000)von Rekowski, R{\"u}diger, \&
  Elstner}]{vRE00}
von Rekowski, M., R{\"u}diger, G., \& Elstner, D. 2000, A\&A, 353, 813

\end{thebibliography}

\end{document}